\documentclass[twocolumn]{aastex61}
\usepackage{txfonts,sans}
\usepackage{graphicx}
\usepackage{natbib}
\usepackage{txfonts}
\usepackage{float}

\def\spb{\smallskip\par\noindent$\bullet\;$}
\received{19 September 2018}
\accepted{30 November 2018}
\submitjournal{AJ}

\shorttitle{Molecular Gas in the Cosmic Eyebrow}
\shortauthors{Dannerbauer et al.}
\begin{document}

\title{Ultra-bright CO and [CI] emission in a lensed $z=2.04$ submillimeter galaxy  with extreme molecular gas properties}

\correspondingauthor{Helmut Dannerbauer}
\email{helmut@iac.es}

\author[0000-0002-0786-7307]{H. Dannerbauer}
\affil{Instituto de Astrof\'{i}sica de Canarias (IAC), E-38205 La Laguna, Tenerife, Spain}
\affiliation{Universidad de La Laguna, Dpto. Astrof\'{i}sica, E-38206 La Laguna, Tenerife, Spain}

\author{K. Harrington}
\affil{Argelander Institute for Astronomy, University of Bonn, Auf dem Hügel 71, D-53121 Bonn, Germany} 
\affil{International Max Planck Research School of Astronomy and Astrophysics at the Universities of Bonn and Cologne}

\author{A. D\'{i}az-S\'{a}nchez}
\affiliation{Departamento F\'{i}sica Aplicada, Universidad Polit\'{e}cnica de Cartagena, Campus Muralla del Mar, E-30202 Cartagena, Murcia, Spain}

\author{S. Iglesias-Groth}
\affil{Instituto de Astrof\'{i}sica de Canarias (IAC), E-38205 La Laguna, Tenerife, Spain}
\affiliation{Universidad de La Laguna, Dpto. Astrof\'{i}sica, E-38206 La Laguna, Tenerife, Spain}

\author{R. Rebolo}
\affil{Instituto de Astrof\'{i}sica de Canarias (IAC), E-38205 La Laguna, Tenerife, Spain}
\affiliation{Universidad de La Laguna, Dpto. Astrof\'{i}sica, E-38206 La Laguna, Tenerife, Spain}
\affil{Consejo Superior de Investigaciones Científicas, E-28006 Madrid, Spain}

\author{R.~T. Genova-Santos}
\affil{Instituto de Astrof\'{i}sica de Canarias (IAC), E-38205 La Laguna, Tenerife, Spain}
\affiliation{Universidad de La Laguna, Dpto. Astrof\'{i}sica, E-38206 La Laguna, Tenerife, Spain}

\author{M. Krips}
\affil{IRAM, Domaine Universitaire, 300 rue de la Piscine, 38406, Saint-Martin-d'H\'{e}res, France}



\begin{abstract}
We report the very bright detection of cold molecular gas with the IRAM NOEMA interferometer of the strongly lensed source {\it WISE} J132934.18$+$224327.3  at $z=2.04$, the so-called Cosmic Eyebrow. This source has a similar spectral energy distribution from optical-mid/IR to submm/radio but significantly higher fluxes than the well-known lensed SMG SMMJ 2135, the Cosmic Eyelash at $z=2.3$. The interferometric observations identify unambiguously the location of the molecular line emission in two components, component CO32-A with I$_\mathrm{CO(3-2)}=52.2\pm0.9$~Jy~km~s$^{-1}$ and component CO32-B with I$_\mathrm{CO(3-2)}=15.7\pm0.7$~Jy~km~s$^{-1}$. Thus, our NOEMA observations of the CO(3-2) transition confirm the SMG-nature of {\it WISE} J132934.18$+$224327.3, resulting in the brightest CO(3-2) detection ever of a SMG. In addition, we present follow-up observations of the brighter component with the Green Bank Telescope (CO(1-0) transition) and IRAM~30m telescope (CO(4-3) and [CI](1-0) transitions). The star-formation efficiency of $\sim$100~L$_\mathrm{\sun}$/(K~km~s$^{-1}$~pc$^{2}$) is at the overlap region between merger-triggered and disk-like star-formation activity and the lowest seen for lensed dusty star-forming galaxies.  The determined gas depletion time $\sim$60~Myr, intrinsic infrared star-formation SFR$_\mathrm{IR}\approx2000$~M$_{\odot}$~yr$^{-1}$ and gas fraction M$_\mathrm{mol}$/M$_\mathrm{\ast}=0.44$ indicates a starburst/merger triggered star-formation. The obtained data of the cold ISM --- from CO(1-0) and dust continuum --- indicates a gas mass $\mu$M$_\mathrm{mol}\sim15\times~10^{11}$~M$_\mathrm{\sun}$ for component CO32-A. Its unseen brightness offers the opportunity to establish the Cosmic Eyebrow as a new reference source at $z=2$ for galaxy evolution.
\end{abstract}




\section{Introduction} 
\label{sec:intro}
The number counts of dusty star-forming galaxies \citep[DSFGs; see for a review][]{cas14}  are steep \citep{bla96,neg07}, thus bright, luminous sources should be rare in the sky. To discover them large infrared surveys covering several hundred to thousand square degrees are needed. The advent of telescopes like {\it Herschel} \citep{pil10}, {\it Planck} \citep{tau10,pla11} and the South Pole Telescope \citep[SPT][]{car11} and their subsequent surveys of large parts of the sky \citep[e.g.,][]{car11,eal10,oli12} provide the indispensable dataset to search for such sources. 

The combination of an increased probability of lensing towards higher redshift \citep[e.g.,][]{bet15}  and the steep number counts of DSFGs facilitate the search and selection of extremely bright star-forming galaxies at the peak epoch of galaxy formation at $z=2$ \citep[e.g.,][]{mad14}. Thus, in the past decade an increasing number of bright, lensed galaxies have been discovered, followed by subsequent observations of the cold molecular gas, the fuel of star-formation \citep[e.g.,][]{neg10,cox11,vie13,can15,har16,har18,yan17}.

Observations of the cold interstellar medium, molecular gas and dust, enables us to study in detail important properties of star-formation processes such as the content of the molecular gas, the star-formation efficiency (SFE) and gas fraction \citep[e.g.,][]{car13}. In case of detecting several CO transitions, the so-called CO spectral line energy distribution (CO SLED), can be constructed and molecular gas properties such as gas excitation and temperature can be studied \citep[e.g.,][]{dan09,dad15}.  

%fig 1  -- integrated map
\begin{figure}[!h] 
\centering
\includegraphics[width=7cm,angle=0]{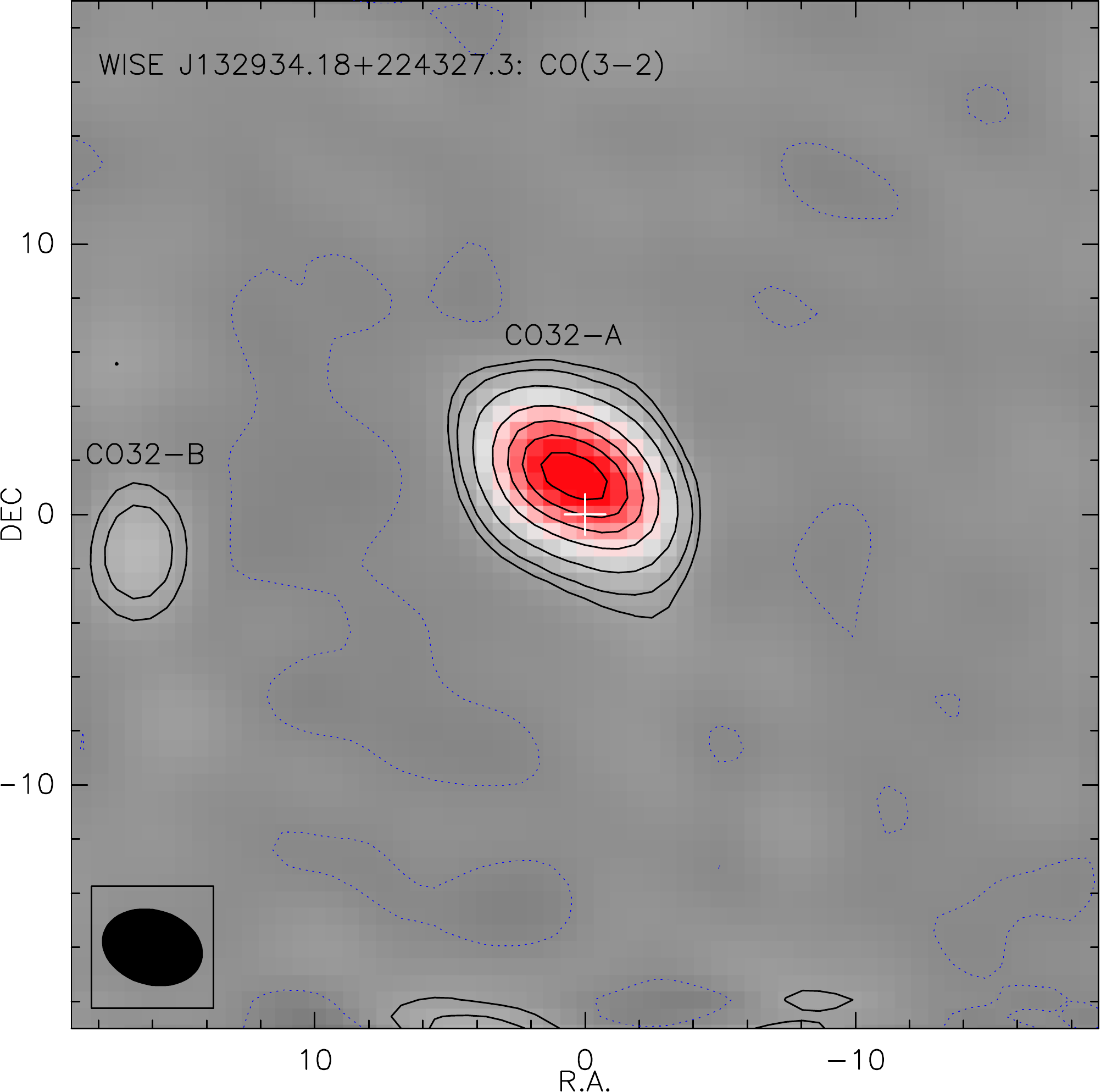}
\includegraphics[width=7cm,angle=0]{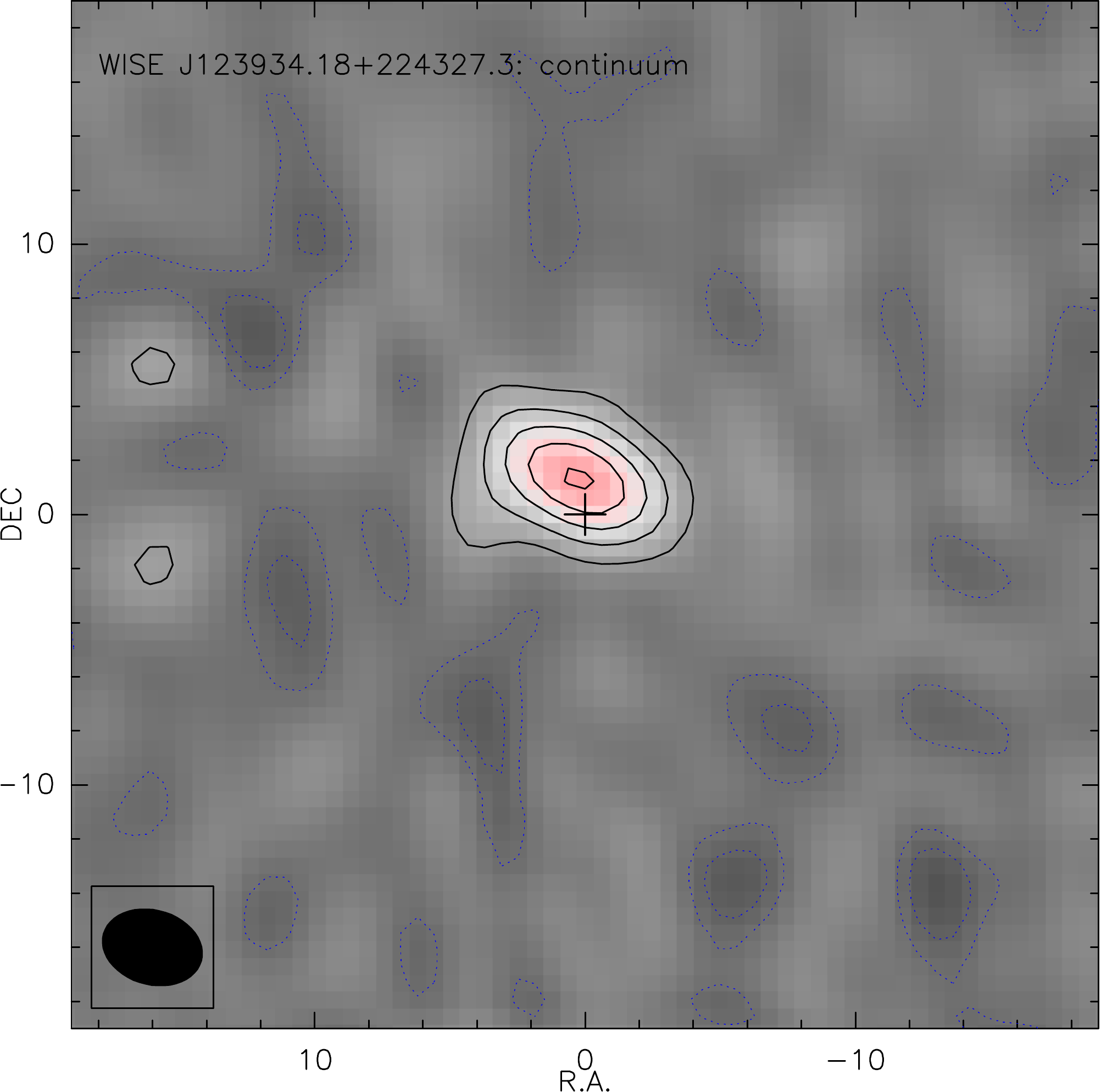}
\caption{Top panel: velocity integrated "cleaned" maps of the CO(3-2) line. The applied velocity range is based on the CO(3-2) line of component CO32-A. Contour levels are: $-$1$\sigma$, 3$\sigma$, 5$\sigma$, 10$\sigma$, 20$\sigma$, 30$\sigma$, 40$\sigma$ and 50$\sigma$.  Bottom panel: dust continuum map averaged over remaining channels. Contour levels are: $-$2$\sigma$, $-$1$\sigma$, 3$\sigma$, 6$\sigma$, 9$\sigma$, 12$\sigma$ and 15$\sigma$. In both maps, the cross is the phase center and the size is $\approx$40$^{\prime\prime}$$\times$40$^{\prime\prime}$. We detect extended emission of the dust continuum at a marginal level.  Higher resolution imaging is indispensable to confirm this possible feature.
}
\label{fig:noemamaps}
\end{figure}

Among the brightest, lensed submillimeter galaxies (SMGs), SMM J2135, the so-called Cosmic Eyelash, at $z=2.3$ \citep{swi10} provides a good reference to further identify in the sky even brighter SMGs which may facilitate subsequent detailed spatial and spectral studies. We carried out a search for bright analogues of similar colours using the VISTA Hemisphere Survey (VHS) \citep{mcm13} and WISE \citep{wri10} over a region of more than 6230 square degress. Details of our source selection and first results of this correlation were published in \citet{igl17}. Subsequently, modifying and extending our technique, \citet{dia17} have shown that cross-matching between the AllWISE and {\it Planck} full-sky compact source catalogues adopting appropriate colour criteria can lead to the identification of extremely bright, lensed SMGs. 

The most promising candidate found in this full-sky search is  {\it WISE} J132934.18$+$224327.3 (alias P1329$+$2243), the so-called Cosmic Eyebrow. This source is found in a strong lensing cluster at $z=0.44$ \citep[][]{ogu12} which has been observed by \citet{jon15} with JCMT/ SCUBA-2, reporting the discovery in snapshot observations of one submillimeter source at 850 and 450 $\mu$m  with S$_\mathrm{450~\mu m}\approx605$~mJy and S$_\mathrm{850~\mu m}\approx130$~mJy \citep{jon15},  and consistent with this source being the main  counterpart of our measured {\it Planck} fluxes.  For comparison, the flux densities of the Cosmic Eyelash at the same wavelengths are S$_\mathrm{450~\mu m}\approx480$~mJy and S$_\mathrm{850~\mu m}\approx115$~mJy \citep{ivi10}.  Within 1 arcsec of the position of the SCUBA-2 submillimeter source, we find in the HST-ACS images a lensed, arc-like galaxy, split into two sources. Low-resolution rest-frame UV-optical spectroscopy of this lensed galaxy obtained with the 10.4~m GTC revealed the typical absorption lines of a starburst galaxy at $z=2.0448\pm0.0004$ \citep{dia17}. Archival Gemini-N near-IR spectroscopy provided a clear detection of H$_{\alpha}$ emission at  $z=2.0439\pm0.0006$. We determined an intrinsic rest-frame 8$-$1000 $\mu$m luminosity, L$_\mathrm{IR}$ of (1.3$\pm$0.1)$\times$10$^{13}$ L$_\mathrm{\odot}$ and a likely star-formation rate (SFR) of $\sim$2000 M$_\mathrm{\odot}$~yr$^{-1}$, taking into account a lensing amplification factor of $11\pm2$ calculated with Lenstool \citep{kne93,jul07}.   At all frequencies from the optical to the radio, the spectral energy distribution (SED) of P1329$+$2243 shows a remarkable similarity to the Cosmic Eyelash but is brighter (up to a factor 4) than the Cosmic Eyelash, and thus it is one of the brightest high-z lensed SMGs ever detected. In this paper, we present our cold molecular gas follow-up of {\it WISE} J132934.18$+$224327.3 with the NOEMA interferometer and two single dish telescopes,  the Green Bank Telescope (GBT) and the IRAM~30m telescope. In section \ref{sec:obs}, we present the observations and in section \ref{sec:results} the results. In section \ref{sec:discussion}, we discuss the properties of our target and conclude in section \ref{sec:conclusion}. We adopt a flat $\Lambda$CDM cosmology from \citet{pla14} with H$_{0}=68$~km~s$^{-1}$~Mpc$^{-1}$, $\Omega_{m}=0.31$,  $\Omega_{\Lambda}=1-\Omega_{m}$,.

%fig 2  -- integrated spectrum of CEB
\begin{figure}[!h] 
\centering
\includegraphics[width=9cm,angle=0]{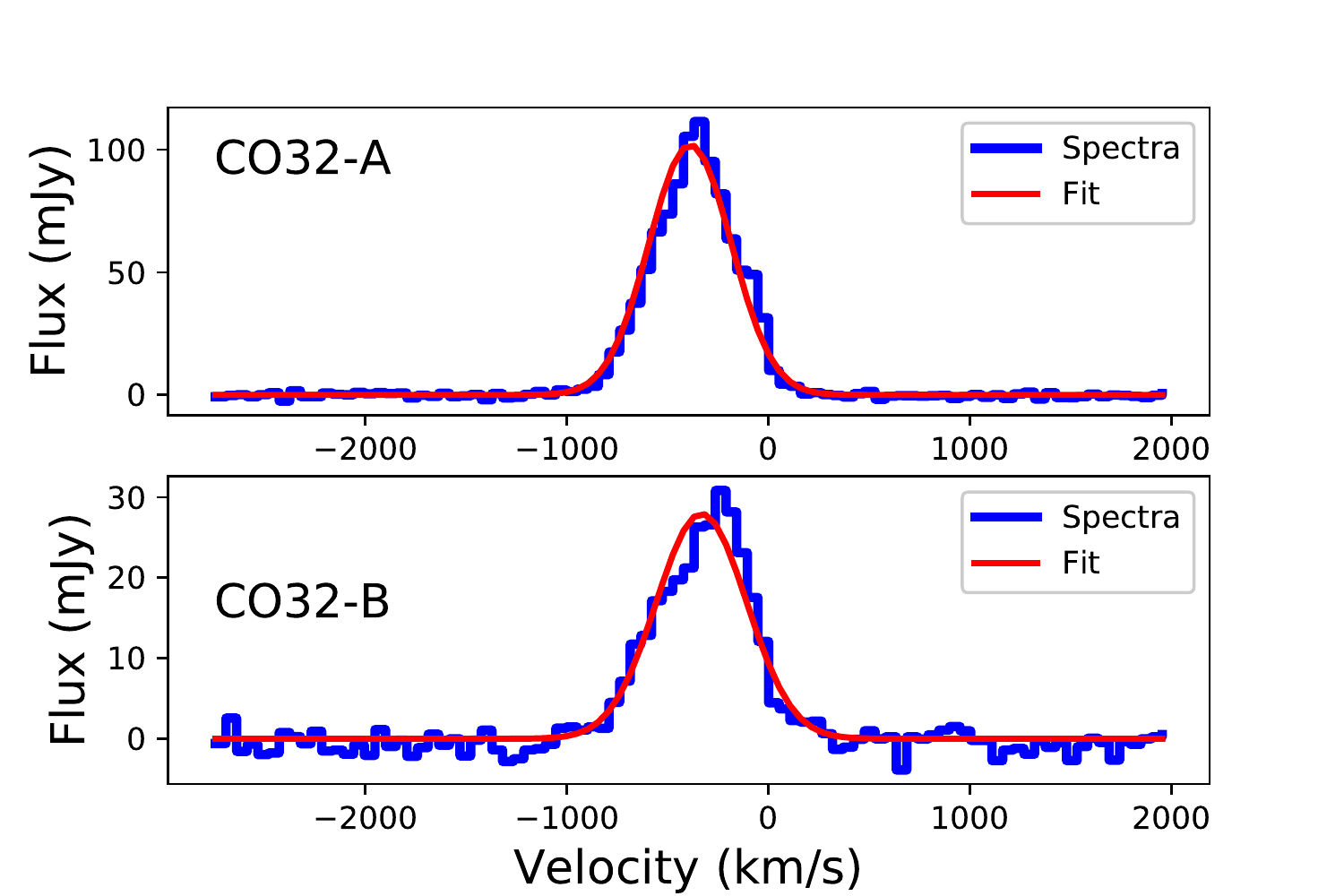}
\caption{CO(3-2) spectra of both images of the Cosmic Eyebrow --- CO32-A at the top and CO32-B at the bottom --- with channel width of 50~km~s$^{-1}$ ($\sim$20~MHz). The red line shows a gaussian fit from which we derived $z_{CO(3-2)}$, $L^{\prime}€™_{\rm CO(3-2)}$ and FWHM$_{\rm CO(3-2)}$. Zero velocity is based on the optical/near-infrared determined redshift $z=2.0439$ \citep{dia17}.}
\label{fig:co1dspec}
\end{figure}

%fig 3 --Overlay CO(3-2) and Continuum on HST image
\begin{figure}[!h]
\centering
\includegraphics[width=8cm,angle=0]{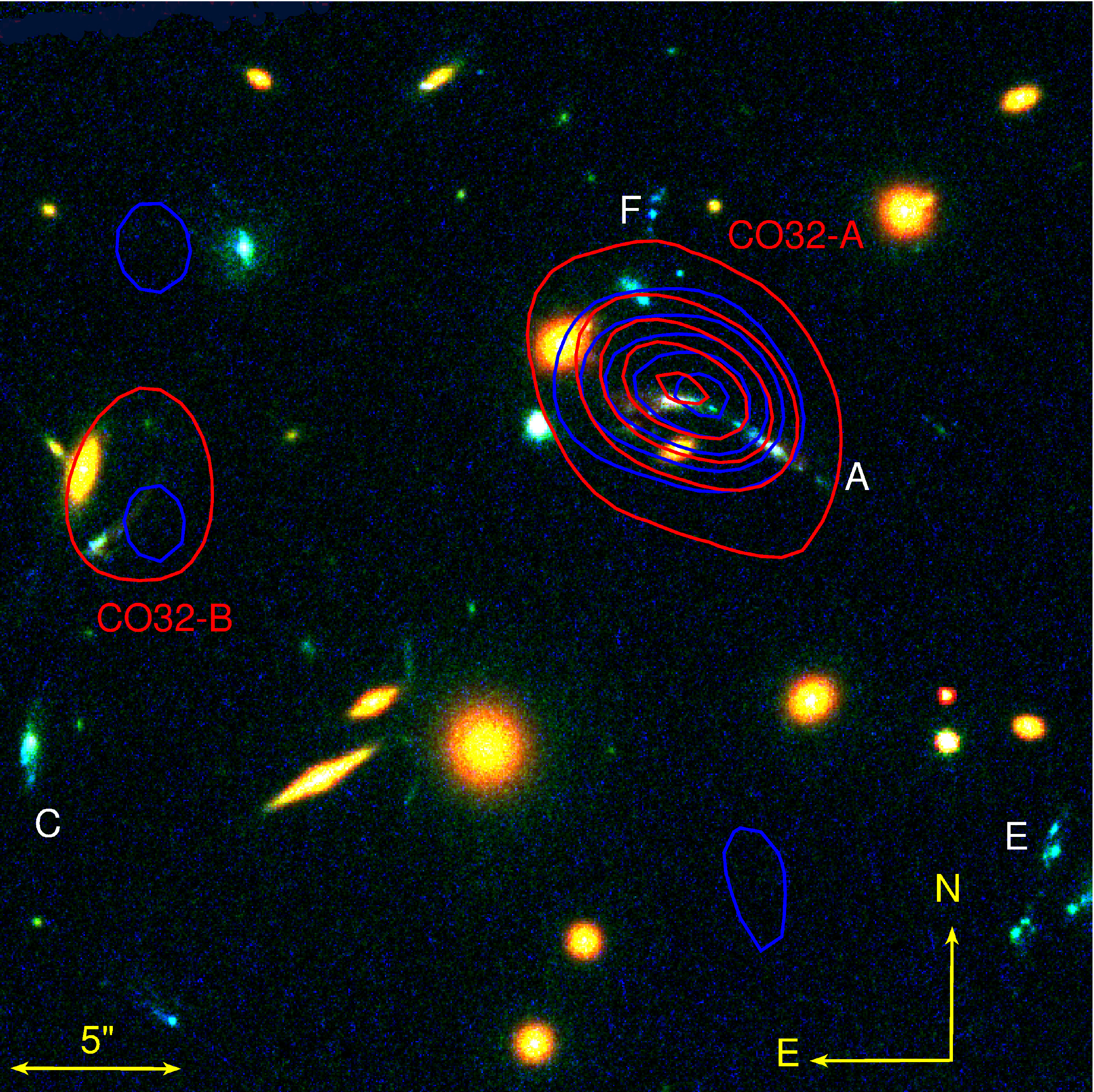}
\caption{The $\approx$$30^{\prime\prime}\times~30^{\prime\prime}$ \textit{HST/WFC3} RGB image (blue: F390W; green: F606W; red: F160W) of the Cosmic Eyebrow and its environment. Several families of multiply lensed background galaxies --- called A, C, E and F \citep[see Fig.~1 in][]{dia17} --- are shown. Total intensity image of the CO(3-2) emission is shown in red contours of the lensed SMG {\it WISE} J132934.18$+$224327.3, contour levels are starting from 3$\sigma,$ and $\sigma$\,=\,0.003 Jy\,beam$^{-1}$ km s$^{-1}$ and then continuing linearly with a spacing of 12.7$\sigma$. In blue contours, the dust continuum at rest-frame 870~$\mu$m is shown, contour levels are starting from 3$\sigma,$ and $\sigma$\,=\,0.004 mJy then continuing linearly with a spacing of 3.3$\sigma$.  }
\label{fig:contours}
\end{figure}

\section{Observations}
\label{sec:obs}
\subsection{NOEMA} 
On 27 August 2017 we observed P1329$+$2243 with IRAM NOEMA for a total of 4.9~hours (DDT D17AA: PI H. Dannerbauer)  with eight antennae in D configuration (E04W08E10N13W12N02W05N09). We used the Wide-X receiver, offering a bandwidth of 3.6 GHz, in dual polarization mode and targeted the redshifted CO(3-2) line at 113.599 GHz (tuning frequency). The phase center is RA$_\mathrm{J2000}=$13:29:34.03 and Dec$_\mathrm{J2000}=+$22:43:25.5. Our observations cover the frequency range from 111.8 to 115.4~GHz, including the CS(7-6) line (342.883~GHz in the rest-frame), expected to lie at 112.790~GHz. The data were calibrated through observations of standard bandpass (3C279), phase/amplitude (1328+307) and flux calibrators (MWC349, 1328$+$307) and reduced with the GILDAS software packages CLIC and MAPPING. Using the deconvolution algorithm/method HOGBOM, we have "cleaned" the "dirty map" of the NOEMA observations.  The FWHM of the synthesized beam is $3\farcs8\times 2\farcs8$ at 113.6 GHz. The field of view respectively the full width at half power of the primary beam (largest scale to be detected) is $44\farcs4\times 44\farcs4$, and includes several families of multiply lensed background galaxies -- called A, C, E and F \citep[see Fig.~1 in][]{dia17} --- used to perform a lens model of the foreground galaxy cluster at $z=0.44$ \citep{dia17}.

\subsection{Green Bank Telescope}
The CO (1-0) line emission was observed using the K$_{a}$ band receiver on the GBT. The pointing position was very close to the NOEMA one (RA$_\mathrm{J2000}=$13:29:34.18 and Dec$_\mathrm{J2000}=+$22:43:27.3). Observations (GBT/17B-305; PI: K. Harrington) took place on October 16th, 2017, under good sky conditions. The FWHM of the main beam at the observed frequency of 37.9~GHz is $\sim$20$^{\prime\prime}$. We used the standard SubBeamNod\footnote{Nodding the 8~m GBT sub-reflector every 6 seconds between each of the two receiver feeds for an integration time of 4 minutes.} procedure, with 4 minute integrations per scan. Pointing and focus were performed before the SubBeamNod integrations, with a follow-up pointing directly to minimize losses in efficiency. The backend spectrometer, VEGAS, was used in its low-resolution 1.5 GHz bandwidth mode tuned to the expected CO (1-0) line frequency. Using GBTIDL \citep{mar13}, we reduced the on-off measurements and corrected for the atmospheric attenuation \citep[see][]{har18}. After dropping bad scans, we smoothed the averaged spectra to 50 km s$^{-1}$ channels, and further removed a low order polynomial outside of the line emission. The resulting on-source integration time was 0.5~hours. Flux  accuracy was checked with the standard source Uranus and pointing stability with 0841$+$7053, 1310$+$3220, 1331$+$3030 and 1642$+$3948. We use a gain conversion factor of 1~K/1.5~Jy and adopt a 25 \% uncertainty for systematic uncertainties  (baseline removal, flux calibration, pointing/focus drifts), and note that the observed frequency lies at the edge of the K$_{a}$ bandpass where receiver performance can be limited.

\subsection{IRAM~30m Telescope}
Using the IRAM 30m, we observed the CO (4-3) and the [CI] (1-0) transitions with the EMIR receiver \citep{car12}, --- same pointing position as for the GBT. The observations (170-17; PI: K. Harrington) took place on February 10th and 11th, 2018 for on-source integrations of 0.75~h and 0.5~h under good weather conditions ($\tau_{225 GHz} = 0.2-0.3$). The FWHM of the primary beam at the observed frequency of 151.7~GHz (observed frequency of the CO(4-3) transition) is $\sim$16$^{\prime\prime}$.  We used the wobbler switching observing mode, with repeated 5 minute integrations consisting of twelve 25 second subscans (wobbler switching frequency of 0.5 Hz; azimuthal offset $=$ 40\arcsec). We recorded the EMIR E150 data with the  fast Fourier Transform Spectrometre (FTS200) backend. Each observing session began with a strong calibration source for pointing and focus measurements. Pointings were  assessed every 1.5 to 2 hours, with azimuth and elevation offsets typically within 2-3\arcsec. The data have been reduced with GILDAS in standard manner and the final spectra binned to $\approx$40~km~s$^{-1}$.

\section{Results}
\label{sec:results}
\subsection{Interferometric Observations with IRAM NOEMA}
\label{sec:resnoema}
We detect a strong signal of the CO(3-2) line at $\sim$113.7~GHz yielding a spectroscopic redshift of $z\approx2.040$. We reveal the molecular gas reservoir via this line at two positions, called CO32-A and CO32-B, separated by $\approx$16\farcs5, see Fig.~\ref{fig:noemamaps} and for details Table~\ref{tab:noema}. 1\farcs5 north away from the phase center, we obtain an impressive $\geq$50 sigma detection of the CO(3-2) line from CO32-A, I$_\mathrm{CO(3-2)}=52.2\pm0.9$~Jy~km~s$^{-1}$ (top panel of Fig.~\ref{fig:co1dspec}), as strong as predicted by us and a factor of four brighter than the Cosmic Eyelash \citep{dan11}. For the second component CO32-B we measure an integrated velocity intensity I$_\mathrm{CO(3-2)}=15.7\pm0.7$~Jy~km~s$^{-1}$ (bottom panel of Fig.~\ref{fig:co1dspec}), taking into account the primary beam correction. We derive FWHMs of 481 (CO32-A) and 529 km~s$^{-1}$ (CO32-B) respectively, typical values for SMGs \citep[e.g.,][]{bot13}. The line profiles are symmetrical. Out of four families of background galaxies \citep[see Fig.~1 in][]{dia17} within the NOEMA field of view, we detect A (CO32-A) and C (CO32-B) (only the member 1 of the C family has CO emission with an offset of $2-3^{\prime\prime}$). In the data cubes we do not see any indication for rotation and outflows of the cold molecular gas in neither of the two components. CO32-A is consistent with the two arc-like rest-frame UV/optical sources 1 and 2 reported in \citet{dia17}, however closer to source 1, see Fig.~\ref{fig:contours}. The second source CO32-B lies 16\farcs5 south-east away from CO32-A. Interestingly, it is consistent with extended emission detected by WISE \citep[see also Fig.~1 in][]{dia17}. In the SCUBA-2 maps at 450 and 850~$\mu$m we see hints for an extended emission towards this second CO source. The peak position of component CO32-B is consistent with an arc for which still no optical spectroscopic redshift exists. The velocity offset between the H$\alpha$ and CO line for component CO32-A is $\sim379$~km/s or $\Delta z=0.0038$ which is consistent with typical offset values of several hundred km~s$^{-1}$ derived from CO \citep{bot13,che17} and H$\alpha$ line observations \citep{swi04,ala12,che17} for SMGs. However, due to the coarse spatial resolution of the NOEMA CO data, we cannot exclude a composite object. We detect the dust continuum at rest-frame 870~$\mu$m for component CO32-A $S_\mathrm{3mm} =1.5\pm0.1$~mJy. The second source CO32-B is marginally seen in continuum (S$_\mathrm{3mm}=0.6\pm0.2$mJy). There is a hint that the dust continuum emission of CO32-A could be extended (Fig.~\ref{fig:noemamaps}). However, due to the coarse beam we cannot measure the morphology of the cold ISM and compare them with the rest-frame UV/optical emission revealed through the {\it HST} imaging shown in Fig.~\ref{fig:contours}. Higher resolution imaging is indispensable to confirm this possible feature. At the end, we note that similar to \citet{dan11} for the Cosmic Eyelash, we did not detect any emission from the faint CS(7-6) line down to 3$\sigma$ (I$_\mathrm{CS(7-6)}<2.7$~Jy~km~s$^{-1}$), deriving a flux ratio CS(7-6)/CO(3-2)$<0.05$, about a factor 2 lower than measured for the Cosmic Eyelash \citep[CS(7-6)/CO(3-2)$<0.12$; ][]{dan11}. 

%fig 4  -- single dish spectra of CEB
\begin{figure}[!h] 
\centering
\includegraphics[width=8cm,angle=0]{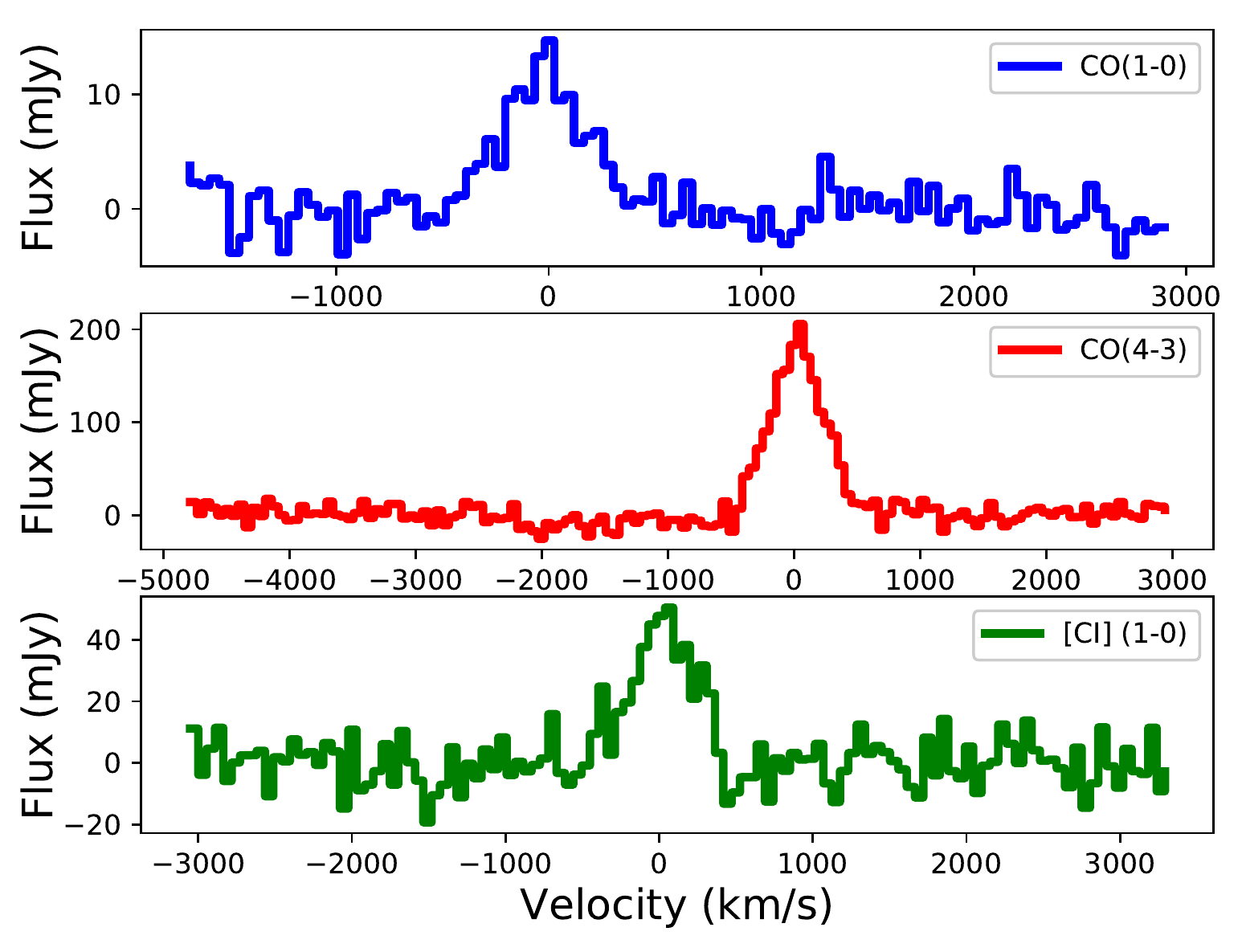}
\caption{We show the single dish spectra obtained for component CO32-A. Top: CO(1-0), middle: CO(4-3), bottom: [CI] (1-0). Zero velocity is based on the redshift obtained with NOEMA.}
\label{fig:co1dsd}
\end{figure}

%==========
%Table1 - NOEMA 
\begin{table*}
\begin{center}
\caption{NOEMA CO(3-2) observations of the Cosmic Eyebrow\label{tab:noema}.}
\begin{tabular}{lcccccccc}
\hline\hline
ID&R.A.&Decl.&z$_{line}$&Frequency&I$_{CO(3-2)}$&FWHM&S$_{CO(3-2)}$&S$_{3~mm}$\\
&(J2000.0)&(J2000.0)&&GHz&Jy~km~s$^{-1}$&km~s$^{-1}$&mJy&mJy\\
\hline
CO32-A &13:29:34.057$\pm$0.001&$+$22:43:26.92$\pm$0.01&$2.04006\pm0.00003$&113.747&52.2$\pm$0.9&481$\pm$7&102.0$\pm$1.2&1.5$\pm$0.1\\
CO32-B &13:29:35.230$\pm$0.004&$+$22:43:24.01$\pm$0.08&$2.04062\pm0.00007$&113.725&15.7$\pm$0.7&529$\pm$17&28.0$\pm$0.8&0.6$\pm$0.2\\
\hline
\end{tabular}
\tablecomments{Units of right ascension are hours, minutes, and seconds, and units of
declination are degrees, arcminutes, and arcseconds.}
\end{center}
\end{table*}
%==========

%==========
%Table2 - Single Dish Observations
\begin{table*}
\begin{center}
\caption{Single Dish Millimeter Observations of the CO32-A\label{tab:gbtemir}.}
\begin{tabular}{llccccccc}
\hline\hline
Telescope&line&z$_{line}$&Frequency&I$_{line}$&FWHM$_{line}$\\
&&&GHz&Jy~km~s$^{-1}$&km~s$^{-1}$\\
\hline
GBT&CO(1-0)&2.04021$\pm$0.00014&37.915&5.7$\pm$1.4&440$\pm$34&\\
IRAM~30m&CO(4-3)&2.04003$\pm$0.00006&151.656&95.4$\pm$14.3&497$\pm$14\\
IRAM~30m&[CI]1-0&2.03994$\pm$0.00019&161.898&21.4$\pm$3.2&450$\pm$41\\
\hline
\end{tabular}
\end{center}
\end{table*}
%==========

\subsection{Single Dish Observations --- GBT and IRAM~30m}
Both with the GBT and IRAM~30m we detect at high significance the targeted lines of CO(1-0), CO(4-3) and [CI] of component CO32-A, see Fig.~\ref{fig:co1dsd} and Table~\ref{tab:gbtemir}. In both observations component CO32-B was not covered by the respective primary beam. We measure integrated velocities for the [CI](1-0)  I$_\mathrm{[CI](1-0)}=21.4\pm3.2$~Jy~K~km$^{-1}$, CO(1-0)  I$_\mathrm{CO(1-0)}=5.7\pm1.4$~Jy~K~km$^{-1}$ and I$_\mathrm{CO(4-3)}=95.4\pm14.3$~Jy~K~km$^{-1}$ transitions. Especially, the CO(4-3) and [CI] emissions are extremely bright, even taking lensing into account, see section \ref{sec:discussion} for a detailed discussion on this. 

Strikingly, the derived spectroscopic redshifts from the different CO transitions are consistent within $\delta z=0.0002$. Thus, the reported velocity offset $\Delta v=379$~km/s (\ref{sec:resnoema}) between the rest-frame optical line H$\alpha$ and the CO(3-2) from the NOEMA observations is real. Furthermore, the profiles and FWHMs of all four cold ISM tracers --- [CI](1-0), CO(1-0), CO(3-2) and CO(4-3) --- are very similar to each other: FWHM~$=440-500$~km~s$^{-1}$.

\section{Discussion}
\label{sec:discussion}
\subsection{General Remarks}
We combine the interferometric and single dish observations to build-up the low-J CO SLED (spectral line energy distirbution) of the Cosmic Eyebrow up to $J=4$. As the GBT and IRAM~30m observations do not cover component CO32-B, we will discuss and compare with the literature the CO SLED of the Cosmic Eyebrow based on component CO32-A. Based on the observations of the cold ISM including several CO transitions, [CI] and dust, we derive physical properties mainly of CO32-A, see Table~\ref{tab:prop}. Whenever possible for completeness we derive physical properties for CO32-B as well, see Table~\ref{tab:prop}. Due to the detection of two components in the far-infrared, we have split and recalculated the infrared luminosity into the two components --- CO32-A and CO32-B. We determined an intrinsic rest-frame 8$-$1000 $\mu$m luminosity, L$_\mathrm{IR}$ of (1.1$\pm$0.1)$\times$10$^{13}$ L$_\mathrm{\odot}$ for CO32-A and L$_\mathrm{IR}$ of (6.8$\pm$0.3)$\times$10$^{12}$ L$_{\odot}$ for CO32-B, following \citet{dia17}.

The CO(3-2) and CO(4-3) transitions are already tracing the star-forming gas. The observed velocity-integrated intensity for these two transitions are extremely high. We will use both transitions in the following section to study in detail the cold ISM of this extremely bright CO source. We stress that due to additional single dish observations with the GBT and the IRAM 30~m telescope, we will estimate the complete cold molecular gas properly, --- via CO(1-0) and [CI](1-0).

\subsection{Lensing}
We constructed the lens model with Lenstool \citep{kne93,jul07} based on the HST ACS F606W image and with only a spectroscopic redshift for family A, see for details \citet{dia17}. Multi-object spectroscopic observations of a large number of possible images of our galaxy and background galaxies are needed in order to construct an accurate lens model. We do not have lens magnification factors derived for the submm continuum nor for the CO(1-0), thus we cannot say anything about potential differential lensing effects \citep{ser12} that may distort this ratio, nevertheless in our preliminary lens model there are spatial magnification gradients in the source plane in the vicinity of the source. 

We can expect that the CO(1-0) will trace even the most diffuse regions \citep[see e.g.,][]{emo16,dan17} of the galaxy, rather than the luminosity-weighted dust temperatures contributing to the observed infrared luminosity. Whereas for the infrared emission a slightly higher magnification is possible if it is on average closer to the caustic. The derived lens model from the optical/NIR data \citep{dia17} and the magnification factors (Table~\ref{tab:prop}) are used for the cold ISM data as well. We are aware that differential lensing could play a role in our galaxy as in the {\it Herschel} and {\it Planck} detected source  HATLAS~J132527$+$284452 at $z=1.68$ \citep{tim15} different magnification factors were derived for the star and dust components. However, our NOEMA data do not have the required resolution (to spatially resolve the source) in order to construct a reasonable lens model. With future high spatial resolution from NOEMA and/or ALMA we should be in a position to verify if differential lensing exists in our target.

\subsection{Are the two CO components images of the same source?}
The lensing magnification factor for the optical-NIR counterparts of CO32-A (sources 1 and 2 in the optical/near-infrared images) is $\mu=11\pm2$. At the position of CO32-B we derive $\mu=15\pm3$. Thus, within the errors the magnification factors at the positions of the CO(3-2) components are consistent. The CO(3-2) line profiles of both CO components of the Cosmic Eyebrow are very similar (Fig.~\ref{fig:co1dspec}). The frequency offset is $\sim$20~MHz which is consistent with the channel width we are using for our analysis. This could suggest that these two CO(3-2) images are from the same background source. However, the optical/near-infrared counterpart of CO32-A is  a factor $\sim$2-3.5 brighter  than CO32-B:  the optical-NIR flux ratio is 2.2, the CO(3-2) is 3.6, the 3~mm continuum is 2.5 and WISE band~1/4 is 2.2/3.1. The magnification factor $\mu$ for the optical/near-infrared counterparts of CO32-B is not smaller than for CO32-A, indicating that the two CO(3-2) components could be different sources. However, between the optical-NIR and CO/mm components of CO32-B, we have an offset of 2-3$^{\prime\prime}$, thus the amplification could be significantly different. High-resolution ALMA imaging is needed to reveal the configuration of this system.

\subsection{Previous CO(3-2) detections of high-redshift sources}
Due to not negligible uncertainties in the conversion of integrated CO fluxes from CO(3-2) into CO(1-0)  \citep[cf.][]{bot13}\footnote{Only a handful of sources in this sample have CO(1-0) measurements.}, we decide to conduct our forthcoming analysis and comparison with the literature with observations of the same transition CO(3-2). Thus, we search the literature for CO(3-2) observations of intermediate- and high-redshift sources. Our literature sample consists of our reference source the Cosmic Eyelash \citep{dan11}, {\it Herschel} \citep{rie13,yan17}, {\it Planck}\footnote{We note that there is an overlap of four sources between the sample from \citet{can15} and \citet{har16}. As the measurements are not consistent in several cases (and no tendencies are seen), we decided to show the measurements from both studies. \citet{can15} use the IRAM~30m telescope and \citet{har16} the LMT for their CO(3-2) measurements.} \citep{can15,har16,har18} and SPT\footnote{In the literature, we only find CO(3-2) observations with measured line flux and FWHM of one SPT-selected source, SPT0311$-$58 at $z=6.900$} \citep{str17} selected lensed SMGs and non-lensed intermediate- and high-redshift ultraluminous infrared galaxies (ULIRGs) \citep{mag14}, normal star-forming galaxies (SFGs) \citep{dad15,ara18} and SMGs \citep{bot13}.

\begin{figure}[th]
\centering
\includegraphics[width=8.0cm,angle=0]{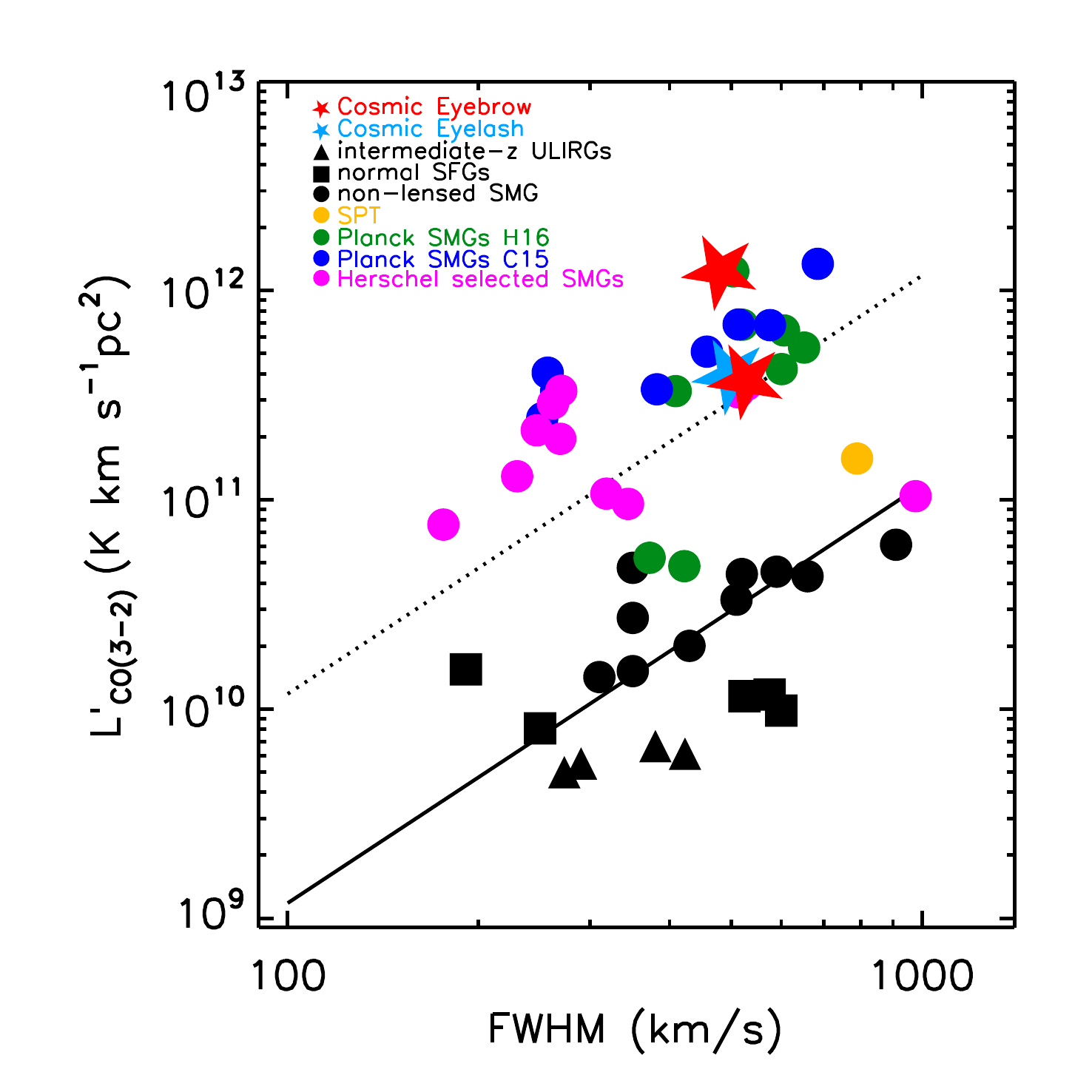}
\caption{Relation between FWHM of the CO(3-2) line and
L$^{\prime}_\mathrm{CO(3-2)}$ high-redshift galaxies. The two CO(3-2) images of the Cosmic Eyebrow are shown as red filled large stars and the Cosmic Eyelash \citep{dan11} as sky blue filled star. The color-coding used for the compilation from literature in the figure is as follows: intermediate-z ULIRGs
\citep[black filled triangle:][]{mag14}, normal SFGs \citep[black filled square:][]{dad15,ara18}, non-lensed SMGs \citep[black filled
circles from][]{bot13}, {\it Herschel}-selected lensed SMGs \citep[magenta filled circles:][]{rie13,yan17}, SPT-selected lensed SMG \citep[gold filled circle:][]{str17}, {\it Planck}-selected SMGs from \citet[][green filled circles]{har16} and \citet[][blue filled circles]{can15}. References for sources can be found in section \ref{sec:discussion}. The solid line shows the relation from \citet{bot13}. The dashed line assumes this relation including a magnification factor of 10.
}
\label{fig:tf}
\end{figure}

\begin{figure}[!th]
\centering
\includegraphics[width=8.0cm,angle=0]{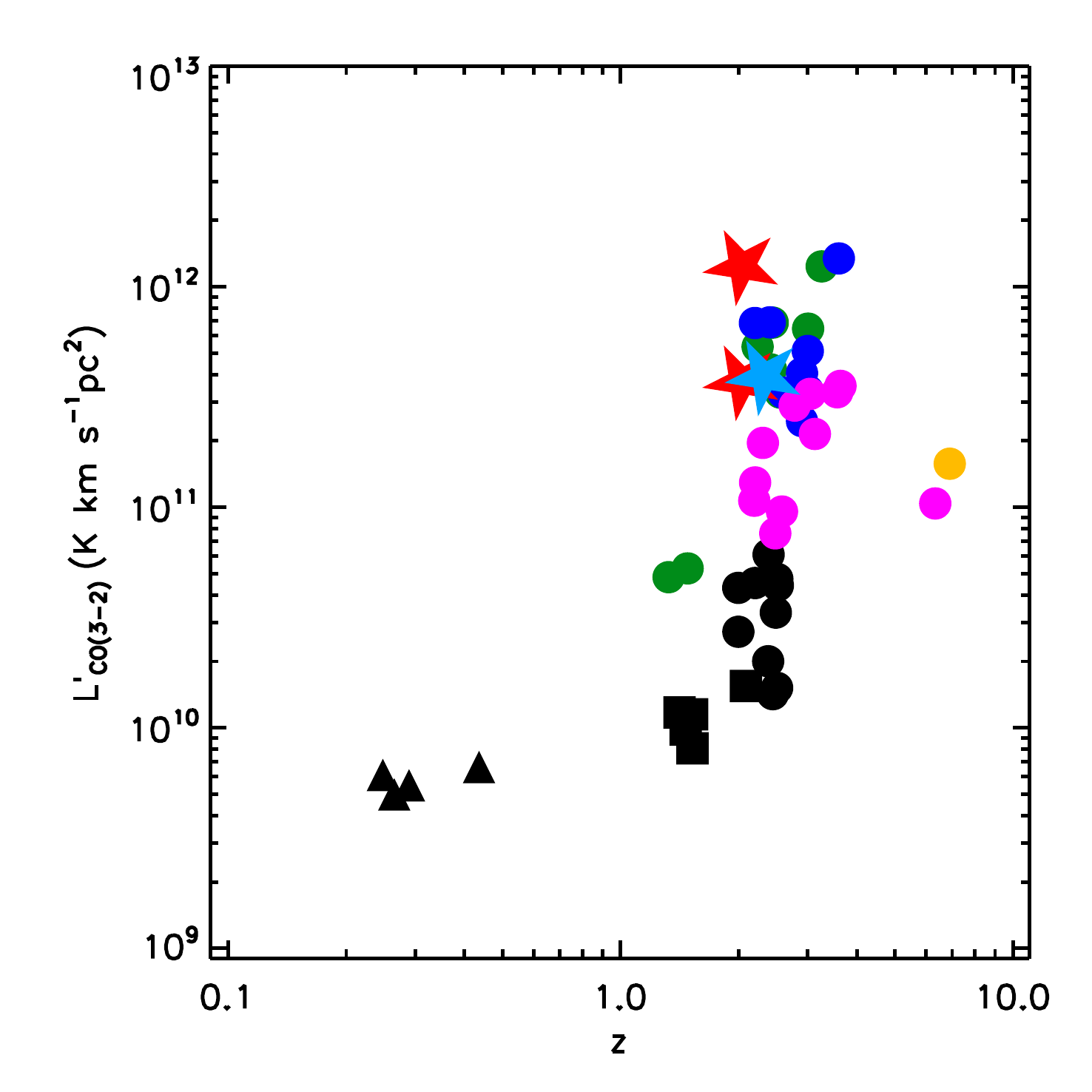}
\caption{
L$^{\prime}_\mathrm{CO}$  as function of redshift. The bright component of the Cosmic Eyebrow is by far the brightest source at redshift $z=2$. At redshifts beyond $z=3.5$ CO(3-2) observations of non-/lensed SMGs are rather sparse (same encoding as in Fig.~\ref{fig:tf}).
}
\label{fig:coz}
\end{figure}

\begin{figure}[!th]
\centering
\includegraphics[width=8.0cm,angle=0]{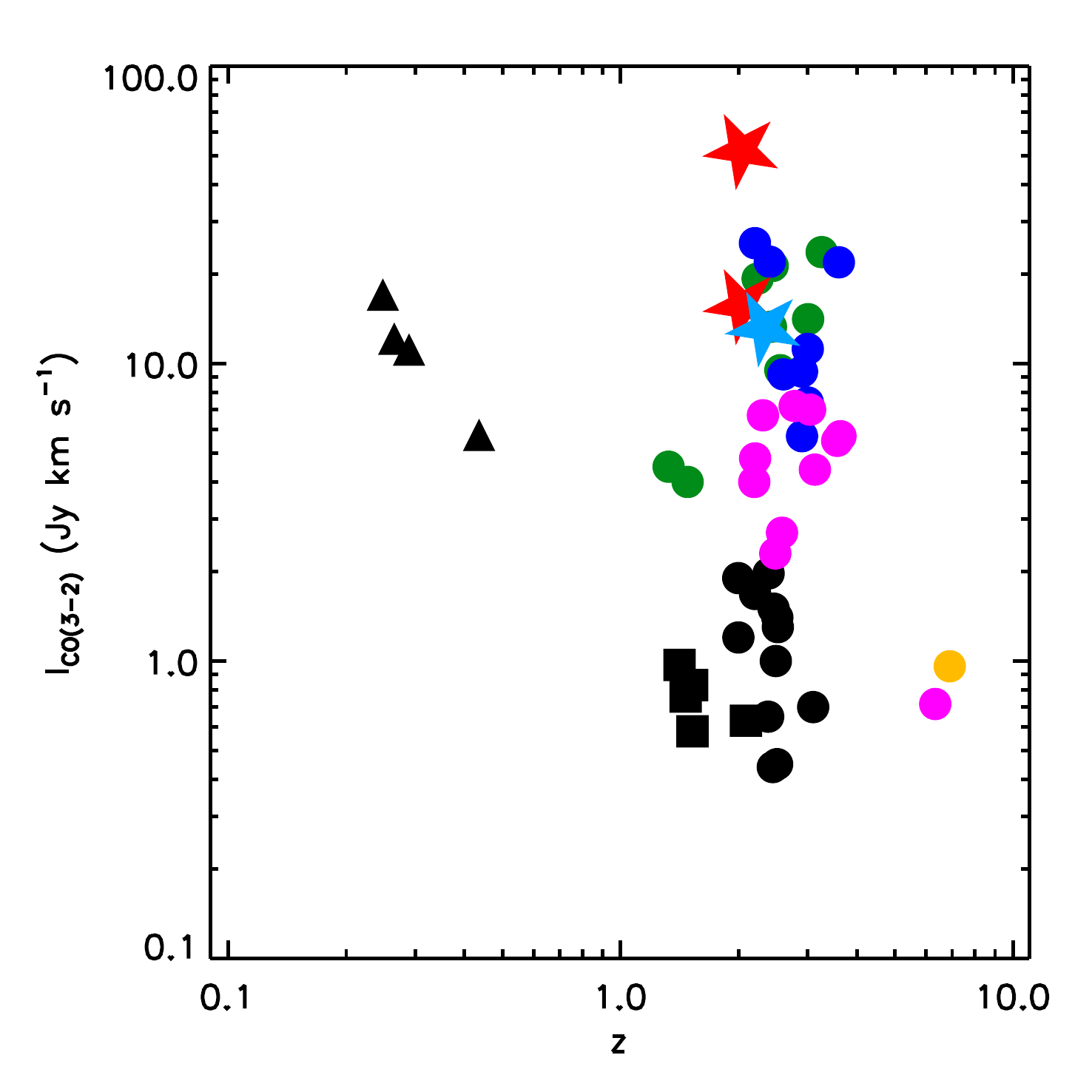}
\caption{
The velocity-integrated intensity I$_\mathrm{CO}$  as function of redshift. The bright component of the Cosmic Eyebrow is by far the brightest source in CO(3-2) in the intermediate- and high-z universe (same encoding as in Fig.~\ref{fig:tf}).
}
\label{fig:icoz}
\end{figure}

The component CO32-A of the Cosmic Eyebrow has the largest velocity-integrated flux ever detected in the CO(3-2) transitions for a high-redshift source (Fig.~\ref{fig:tf}, \ref{fig:coz} and \ref{fig:icoz}). The observed L$^{\prime}_\mathrm{CO}$ is the largest together with two other {\it Planck}-selected lensed SMGs, PJ160917.8 at $z=3.2$ \citep{har18} and PLCK\_G145.2+50.9 at $z=3.6$ \citep{can15}, see Fig.~\ref{fig:tf}. At redshift $z=2$, the peak epoch of star-formation and black hole activity \citep[e.g.,][]{mad14}, the CO(3-2) luminosity of {\it WISE} J132934.18$+$224327.3 is unseen (Fig.~\ref{fig:coz}).

%fig 5 -- L'_CO  vs. FWHM ... a kind of T-F

%fig 6 -- L?CO vs z

%fig 7 -- L?CO vs z

\subsection{Molecular gas mass}
The CO(1-0) line emission is optically thick and can be assumed to trace the spatial extent and bulk of molecular ISM gas mass. Thus, we derive the molecular gas mass directly from the CO(1-0) observations. Although there exists commonly used conversions between higher-J transition and CO(1-0) for SMG \citep{bot13}, the Cosmic Eyebrow is a good example that covering the lowest CO transitions (J$\leq$2) is indispensable in order to obtain an accurate measurement of the total cold molecular gas mass (Fig.~\ref{fig:cosled}). The derived star-formation efficiency of our source (Fig.~\ref{fig:sfe}) suggests to use the CO luminosity to molecular gas-mass conversion factor  $\alpha_\mathrm{CO}=0.8$~M$_{\odot}$~pc$^{-2}$ (K~km~s$^{-1})^{-1}$, a commonly used value for merger-induced star-formation \citep[][]{sol05}. Including a factor of 1.36 to account for Helium, we derive $\mu$M$_\mathrm{mol}=13.2\pm 3.3 \times10^{11}$~M$_\mathrm{\odot}$. We note that using the CO(3-2) velocity-integrated flux density, the molecular gas mass would have been overestimated by a factor of 2. The brightness ratio for SMGs is r$_{32}=0.52$ \citep{bot13}, however the Cosmic Eyebrow is thermalized yielding to  r$_{32}=1.01$ (Table~\ref{tab:prop} and Fig.~\ref{fig:cosled}).

Both carbon fine structure emission lines, [CI](1-0) and [CI](2-1), have been proposed as molecular gas mass tracers \citep[e.g.,][]{wei03,wei05,pap04}  due to their low-excitation requirements (23.6 K and 62.5 K energies above ground state). Neutral carbon is easily influenced by turbulent mixing, such that the relative neutral carbon distribution remains constant in various ISM environments \citep{xie95}. 

The abundance is often assumed in other studies when lacking low-J CO measurements in order to derive the M(H$_{\rm 2}$) \citet[e.g.,][]{wei03} via the atomic to molecular weight ratio of M(CI)/$6\times$M(H$_{\rm 2}$). Here we derive the X(CI)/X(H$_{\rm 2}$) abundance by first using Eq. 2 from \citet{wei05} to compute the total neutral carbon gas mass $\mu$M$_{\rm [CI] } = 3.3 \times 10^{8}$ M$_{\odot}$, assuming a neutral carbon excitation temperature of T$_{\rm exc} = 35$ K. The excitation temperature only changes the total carbon mass significantly when T$_{\rm exc} < 20$ K \citep[see][]{wei05}. We assume the upper limit carbon excitation temperature,  T$_{\rm exc}=35$~K, in the sample of QSO/SMGs from \citet{wal11} because of the extreme nature of the Cosmic Eyebrow. The velocity-integrated [CI](1-0) flux of roughly 21 Jy km~s$^{-1}$ is 5$\times$  that of the strongly lensed Cloverleaf QSO \citep{wei05}, which is amongst the brightest on average in the high-z universe \citep{wal11,ala13,bot17}. The thermalized low-J CO line ratios suggests the higher value of 35 K is appropriate to assume in this system. The independent CO(1-0) derived H$_{\rm 2}$ mass (assuming an $\alpha_{\rm CO} =$ 0.8), provides an abundance of M$_{\rm CI} / 6\times$ M$_{\rm H_{2}} = 5.4 \times 10^{-5}$ --- which is comparable to the value of 4$\times 10^{-5}$ found in nearby galaxies, and within the lower value found in the average carbon abundance of X([CI])/X(H$_{\rm 2}$)= 8.5 $\pm$ 3.5 in \citet{wal11}. This implies a strong carbon enrichment in this system \citep{wei05} at $z=2.04$.

Using the SCUBA-2 measurement of S$_\mathrm{850~\mu m}=127$~mJy ($\lambda_{\rm rest}=250~\micron$) from \citet{jon15} including both CO components, and our NOEMA 3~mm continuum measurement of 1.5 mJy ($\lambda_{\rm rest}=870~\micron$) of CO32-A, we are able to derive a range for the molecular ISM mass using the empirical calibration of long wavelength thermal dust emission to total gas mass, as presented in \citet{sco14,sco16,sco17}. We adopt a dust temperature, T$_{\rm d}=$~T$_\mathrm{ex}=35 $K, which is 3 K less than the luminosity-weighted dust temperature derived in the SED fit to the Cosmic Eyelash template \citep{dia17}. The assumed colder dust temperature will trace more of the mass and will be easier to compare to the [CI](1-0) and CO(1-0) derived gas masses. T$_{\rm d}=35$~K is equal to the highest reasonable mass-weighted dust temperature suggested in \citet{sco16}. We would not expect much lower dust temperatures given an intrinsic far-IR luminosity more than log(L$_{IR}$/L$_{\odot}$)~$\sim13$ \citep[magnification factor $\mu\sim11$;][]{dia17}. 

With fixed T$_\mathrm{d}$ and the observed dust continuum, we use Eq. 16 in Scoville 2016 to derive $\mu$M$_\mathrm{mol}=30.9\pm6.2\times10^{11}$~M$_{\odot}$ from the observed 850 $\micron$ for the total Cosmic Eyebrow system, and $\mu$M$_\mathrm{mol}=15.1\pm~3.0\times10^{11}$ M$_{\odot}$ for component CO32-A from the observed 3~mm. Assuming the IR-luminosity ratio of 2.8 between CO32-A and CO32-B to separate the SCUBA-2 flux at 850~$\mu$m, see also Table~\ref{tab:prop}, we obtain $\mu$M$_{mol}=22.8\pm4.6\times10^{11}$~M$_{\odot}$ from the observed 850 $\micron$. The empirically calibrated equation holds for $\lambda_{\rm rest} \ge 250~\micron$. As the rest wavelength from the observed 850 $\micron$ measurement is almost exactly $250~\micron$, our 3mm continuum measurement can be used to exclusively trace the total ISM mass via longer-wavelength thermal dust emission. We note that the restricted stellar mass range, M$_{\star} = 2\times10^{10} - 4\times10^{11}  $ M$_{\odot}$,  used to calibrate the long-wavelength dust continuum to total gas mass is appropriate, as our MAGPHYS \citep{dac12} derived stellar mass \citep[for the optical/near-infrared counterparts source 1 and 2 in][]{dia17} corrected for amplification is at the upper limit: M$_{\star} = 3.1\times10^{11}  $ M$_{\odot}$ \citep{dia17}. 

Using only the 850 $\micron$ flux for CO32-A would suggest decreasing the neutral carbon metallicity to a value of roughly X(CI)/X(H$_{2}$)~$\approx3.0\times10^{-5}$ (keeping everything else fixed from our assumptions above). This is unlikely for the expected rapid build up of heavy metals during the starburst phase within the Cosmic Eyebrow. Additionally, the ISM mass derived from only the 850 $\mu$m photometry would suggest an $\alpha_{\rm CO}$$\sim2\times$ higher when comparing to our CO(1-0) derived gas mass. We conclude that there are likely two dominant molecular gas phases (with separate $\alpha_{\rm CO}$), as seen in local ULIRGs \citep[e.g.,][]{liu17}, and that other studies such as multi-J CO and [CI] non-LTE radiative analysis are required for further clarification.

To summarize, most of the gas mass measures suggest a cold molecular gas mass of $\mu$M$_\mathrm{mol}\approx15\times~10^{11}$~M$_\mathrm{\sun}$ for component CO32-A and that applying the  conversion factor $\alpha_{CO}=0.8$ typical for merger induced starburst is reasonable for our source.

\subsection{Properties of the Cosmic Eyebrow}

\subsubsection{Morphology}
The astrometric calibration of the HST data was done with stars from SDSS. The error is 0\farcs07 and the astrometry of the NOEMA dataset is based on the radio frame. The offset between the CO(3-2) position and the rest-frame optical source 1 is 0\farcs21 (see Fig.~\ref{fig:contours}) and within the position uncertainty of 0\farcs3 of the NOEMA detection. Furthermore, the NOEMA synthesized beam at the observed frequency has a size of is $3\farcs8\times2\farcs8$ at 113.6 GHz. Due to the obtained spatial resolution, we cannot establish the geometry of the rest-frame UV/optical neither of the cold molecular gas components, e.g., we cannot rule out that there exists an offset between the millimeter emission and the rest-frame optical emission even from the rest-frame UV/optical source 1 \citep[reported in][]{dia17}.  

At a marginal level we detect extended dust emission for CO32-A (Fig.~\ref{fig:noemamaps}). Confirming this possible feature\footnote{We note that decorrelation on the longer NOEMA baselines may cause this feature as well.} with higher resolution mm-imaging would be a very interesting and surprising result as recent ALMA observations of a few SMGs from the ALESS survey \citep[e.g.,][]{hod13,kar13} show that the dust continuum is about a factor of two to five smaller than the cold molecular gas reservoir \citep{che17,cal18} 

There is no sign of a double-horn rotation feature and in the NOEMA data cube we did not detect any rotation. With the current spatial resolution from the NOEMA observations of about $\sim3^{\prime\prime}$, however, we cannot rule out the possibility of a late stage merger event  where the rotation of both galaxies would blend into one single gaussian \citep[][]{eng10}.

\subsubsection{Star-formation efficiency}
The derived star-formation efficiency SFE$=$SFR/L$^{\prime}_\mathrm{CO(1-0)}$$\approx100$~L$_{\sun}$/(K~km~s$^{-1}$~pc$^{2}$) for the Cosmic Eyebrow is at the lower end expected for high-z, merger-triggered star-formation but the gas depletion of $\sim$60~Myr (Fig.~\ref{fig:sfe}) is consistent with high-z starburst activity. This star-formation mode is strengthened by the gas fraction M$_\mathrm{mol}$/M$_\mathrm{\ast}=0.44$, a value expected for starbursts at redshift $z=2$, see e.g. Fig.~6 in \citet{dan17}. The relation described in \citet{whi12} predicts the star-formation rate of a main-sequence galaxy SFR$_\mathrm{MS}=220$~M$_{\odot}$~yr$^{-1}$, the derived lensing corrected SFR$_\mathrm{IR}=1970$~M$_{\odot}$~yr$^{-1}$ for CO-32A is a factor $\sim9$ higher. 

\subsubsection{CO spectral line energy distribution}
\begin{figure}[ht]
\centering
\includegraphics[width=9cm,angle=0]{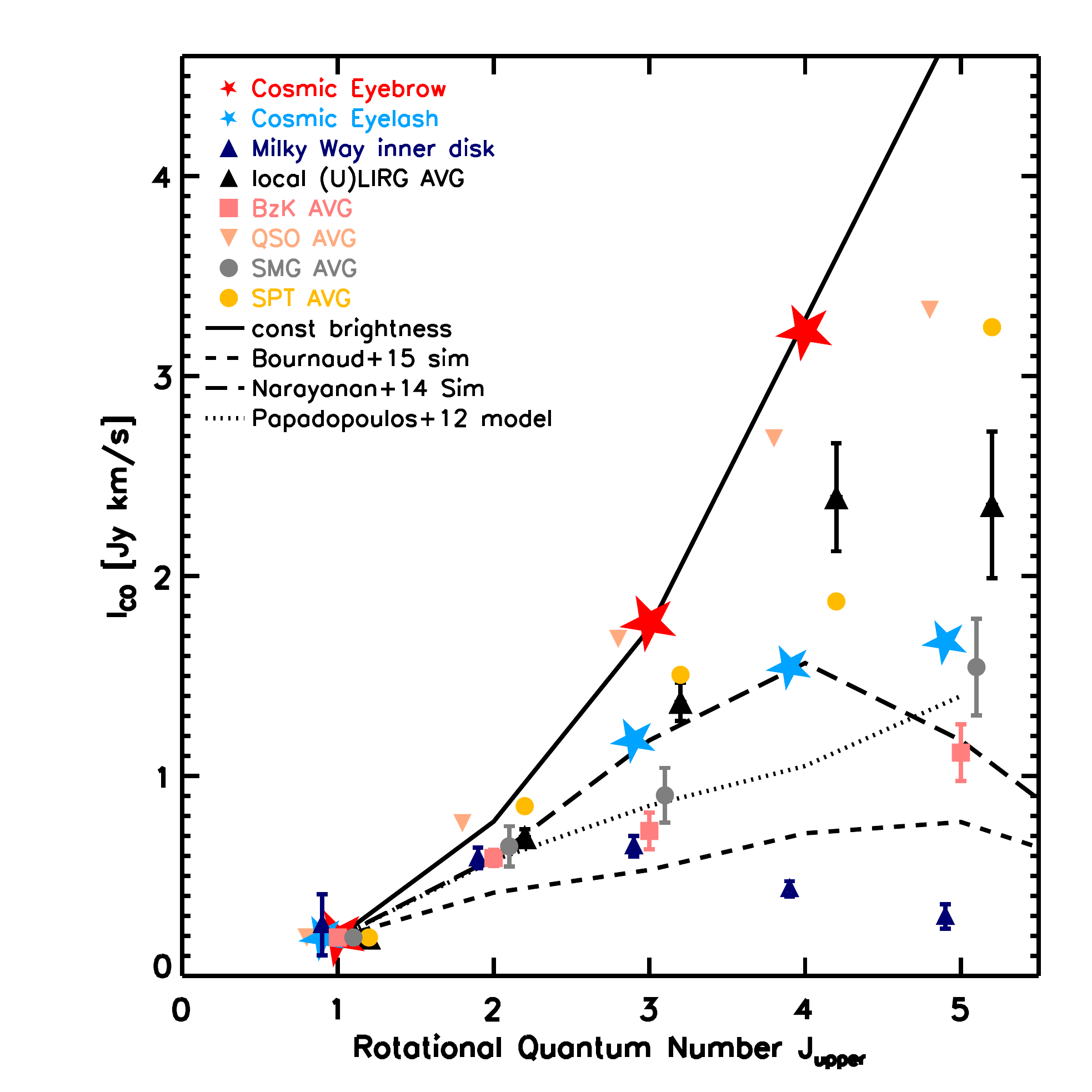}
\caption{Adaption from the compilation of \citet{dad15} showing the CO SLED for different galaxy populations and models/predictions. We compare the Cosmic Eyebrow (large red filled star) with the Cosmic Eyelash (large sky blue) filled star,  average SLED of normal star-forming galaxies \citep{dad15} selecting through the BzK-selection criteria \citep{dad04},  the Milky Way SLED \citep{fix96}, the average of SMGs \citep{bot13}, the average of lensed SMGs selected from the SPT survey \citep{spi14}, the average of QSOs \citep{car13} and the average (U)LIRGs SLED derived in this paper using measurements from \citet{pap12}. All CO SLEDs are normalized to the CO(1-0) of the average BzK galaxy CO SLED, except the Milky Wat which is normalised using CO(2-1). The results from the toy model of \citet{pap12} and the numerical simulations of \citet{bou14} and \citet{nar14} are also shown. Clearly, the CO SLED of the Cosmic Eyebrow stands out and is completely thermaziled up to CO(4-3).}
\label{fig:cosled}
\end{figure}

\begin{figure}[!th]
\centering
\includegraphics[width=8.0cm,angle=0]{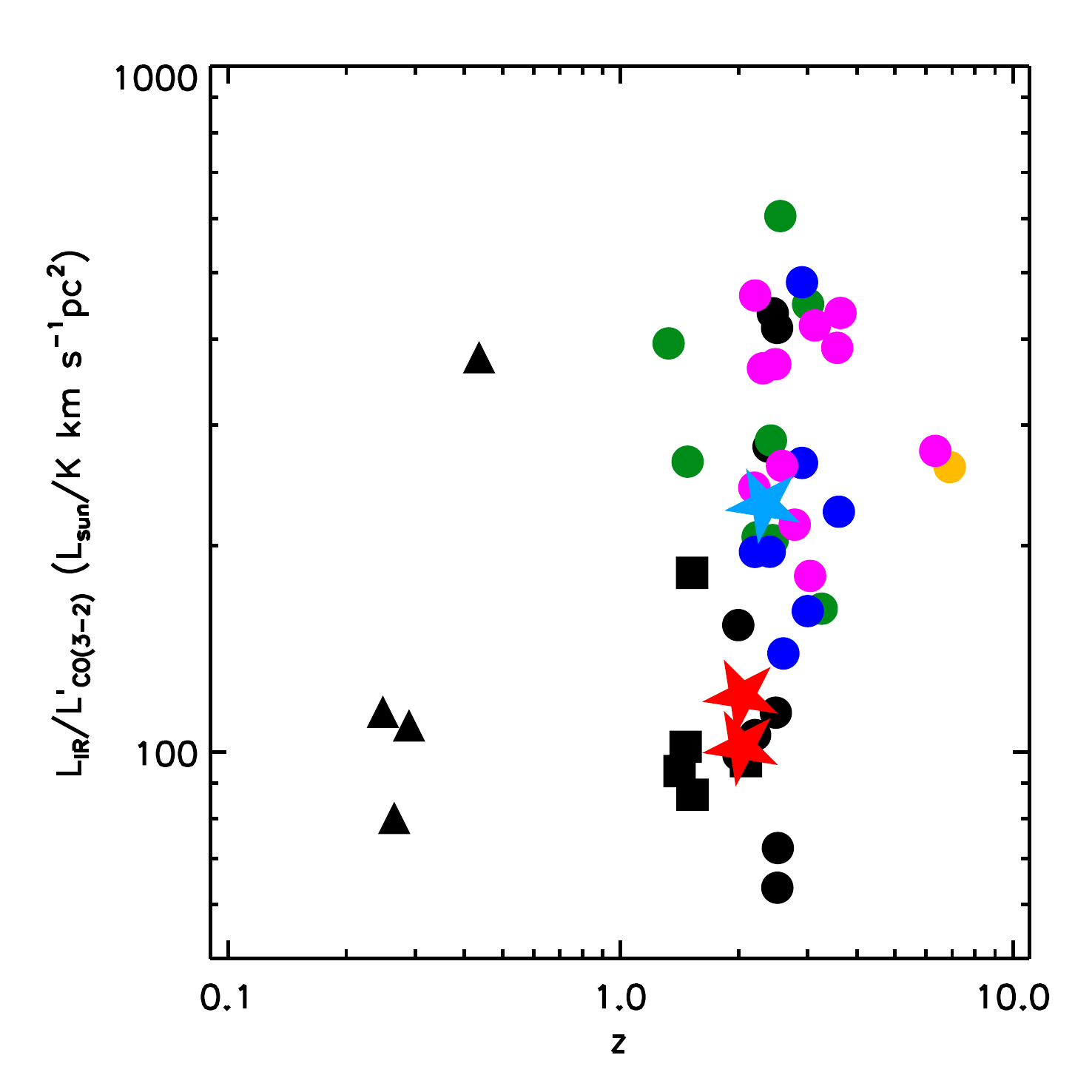}
\caption{
The L$_\mathrm{IR}$-to-L$^{\prime}$$_\mathrm{CO(3-2)}$ ratio as function of redshift. Compared to other lensed SMGs, the SFE of the Cosmic Eyebrow is rather low but consistent with non-lensed SMGs and higher than for normal star-forming galaxies (same encoding as in Fig.~\ref{fig:tf}).
}
\label{fig:sfe}
\end{figure}

The CO SLED of {\it WISE} J132934.18$+$224327.3 is thermalized even up to CO(4-3), see Fig.~\ref{fig:cosled}. It deviates significantly from the Cosmic Eyelash \citep{dan11}, the average non-lensed SMG \citep{bot13}, and the average lensed SMG  \citep[selected from the SPT survey][]{spi14} --- the former both are similar ---, resembling high-z QSOs \citep{car13}. The CO SLED of the Cosmic Eyebrorw. Although the selection is based on the optical to far-infrared SED of the Cosmic Eyelash, a typical SMG, the CO SLED is completely different. Up to CO(3-2), QSOs and hybrid SMGs/QSOs are thermalized as well \citep{sha16}.  Strikingly, the CO(4-3) transition makes the difference. We note that our source is seen in the VLA survey FIRST \citep[S$_\mathrm{1.4~GHz}=3.56\pm0.14$~mJy;][]{bec94} indicating AGN activity. Thus, the radio can help to distingush between AGN and starburst dominated sources. \citet{har18} presented the low-J CO observations of seven {\it Planck} and {\it Herschel} selected sources. None of the sources with $J>2$ shows such a behaviour as P1329$+$2243 although this source is selected from {\it Planck} data as well. PJ160917.8 \citep{har18} has a similar apparent L$^{\prime}_\mathrm{CO}$ as the Cosmic Eyebrow however its CO SLED resembles rather typical SMGs which are dominated by starburst activity. Unfortunately, for lensed SMGs selected from {\it Planck} by \citet{can15} no CO SLEDs are published, especially the CO(1-0) transition is missing.
%-----------------------
%fig 8 -- CO SLED

\subsubsection{[CI] and CO brightness temperature relations}
We derive L$^{\prime}_\mathrm{[CI(1-0)]}$/L$^{\prime}_\mathrm{CO(1-0)}=0.22$. This value is consistent within the lower bound of the average value of L$^{\prime}_\mathrm{[CI(1-0)]}$/L$^{\prime}_\mathrm{CO(1-0)}=0.29\pm0.12$ derived for high-z sources such as SMGs and QSOs in \citet{wal11}. It is closer to the value of L$^{\prime}_\mathrm{[CI(1-0)]}$/L$^{\prime}_\mathrm{CO(1-0)}=0.2\pm0.2$ found in local and nearby galactic systems \citep{ger00}, while the [CI] to CO(1-0) ratio is higher than in the Milky Way 
\citep[0.15$\pm$0.1;][]{fix99}. The derived L$^{\prime}_\mathrm{[CI(1-0)]}$/L$^{\prime}_\mathrm{CO(3-2)}=0.22$ is comparable to the value 0.32$\pm$0.13 for high-z star-forming systems \citep{wal11}.

Including the CO(4-3) transition to the CO(1-0) and [CI](1-0) detections enables us to get more insight into the excitation conditions of the cold ISM of our source. We compare our source with low and high redshift sources from the literature \citep{dan11,les10,les11,isr15,kam16,ros15,emo18}. As the molecular gas is thermalized at least to the CO(4-3) transition,  we would expect a significant component of warmer gas to be associated with the low-excitation gas presented in this study.

\begin{figure}[!th]
\centering
\includegraphics[width=9.0cm,angle=0]{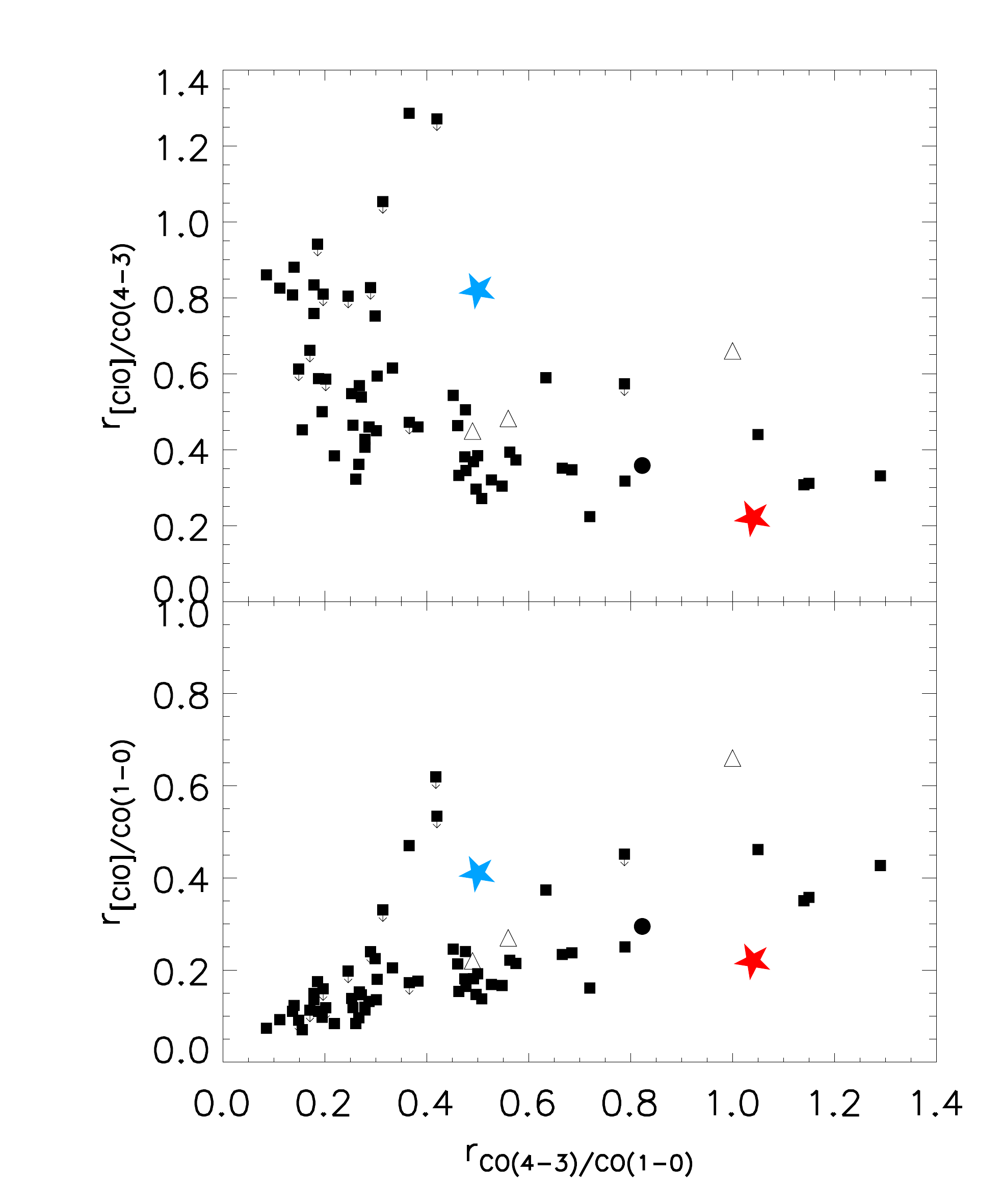}
\caption{Adaption from the compilation of \citet{emo18}. Brightness temperature relations between [CI], CO(1-0) and CO(4-3) are shown. The Cosmic Eyebrow is shown as red filled large star and the Comic Eyelash \citep{dan11} as sky blue filled star. The lensed DSFG MM18423$+$5938 at $z=3.93$ is shown as black filled circle \citep{les10,les11}, low-z star-forming galaxies are shown as black filled squares \citep{isr15,kam16,ros15} and several members of the Spiderweb protocluster MRC1138$-$262 at $z=2.16$ with detections in both CO lines are shown as black open triangles \citep{emo18}.}
\label{fig:ism}
\end{figure}

In Fig. \ref{fig:ism} we show that there is an extreme amount of CO luminosity compared to the [CI]. For such a short-lived starburst episode we expect that the gas will no longer be thermalized after the burst of star formation when conditions will be less intense. The r$_{41}$  --- brightness temperature ratio of the CO(4-3) and CO(1-0) transitions) --- will decrease from unity and the [CI]/CO ratio will likely evolve towards the more populated parameter space once the gas supply is consumed.

%fig 9 -- SFE vs z

%fig 10 -- CI, CO4-3, CO1-0 plot

\subsubsection{CO line width-luminosity relation}
We explore the relation between FWHM and L$^{\prime}_\mathrm{CO}$ of the CO(3-2) line for our compiled sample from the literature. This relation was first proposed by \citet{bot13} for SMGs and then extended to normal SFGs at high-z in \citet{dan17}. This relation can be used as an indicator to see if a high-z CO bright source could be strongly lensed or not \citep{har12,ara16,yan17,har18}. In Fig.~\ref{fig:tf}, a segregation between lensed and non-lensed sources is clearly seen. Due to the dispersion of the relation between FWHM and L$^{\prime}_\mathrm{CO}$, it is not possible to calculate even approximate magnification factors as suggested previously by \citet{har12} and \citet{ara16}. However, from  Fig.~\ref{fig:tf} it is clear that our source has a strong magnification as predicted by our lens model.

\section{Conclusion}
\label{sec:conclusion}
In this work we present a detailed study of the cold ISM with IRAM NOEMA, GBT, and IRAM 30~m of an ultra-bright, lensed submillimeter galaxy at $z=2.04$ with extreme molecular gas properties. The main results are the following:

\spb We have revealed the location of the SMG  {\it WISE} J132934.18$+$224327.3 at $z=2.04$ with IRAM NOEMA via CO(3-2) observations unambiguously and find two components, CO32-A and CO32-B. The derived flux ratios and magnification factors do not discard a two-component/merger-like system, however with the current low spatial resolution dataset in the mm-regime and the preliminary lens model,  we cannot exclude that these two images are of the same source.
\spb The determined redshifts and FWHMs of the four cold ISM line tracers [CI], CO(1-0), CO(3-2) and CO(4-3) are very similar.
\spb In combination with single-dish observations from the GBT and IRAM 30~m telescope we have built up the CO SLED of component CO32-A of the Cosmic Eyebrow and show that the molecular gas is thermalized up to the CO(4-3) transition.
\spb Based on the results derived from the CO SLED, we emphasize that to constrain the cold molecular gas mass, observations of low-J transitions are indispensable. Using suggested conversion for SMGs from e.g., CO(3-2) into CO(1-0) should be used with great caution.
\spb We emphasize that the measured integrated velocity intensities for the CO(3-2), CO(4-3) and [CI(1-0)] transitions have been unseen so far in the early universe. 
\spb Our dataset enables us to measure the cold ISM of the Cosmic Eyebrow via two measures: CO(1-0) and dust continuum. Overall both methods suggest a cold molecular gas mass $\mu$M$_\mathrm{mol}\sim15\times~10^{11}$~M$_\mathrm{\sun}$.
\spb The gas depletion time of $\sim$~60~Myr suggest a merger induced star-formation, strengthened by the measured infrared SFR and gas fraction.

The very high apparent flux brightness offers the opportunity to get new insights in the star-formation processes of high-z galaxies at the peak epoch of the star-formation and black hole activity in the universe. The Cosmic Eyebrow may become a new reference source at $z=2$ for galaxy evolution. Although the observed luminosity is exceptional large, only two other {\it Planck}-selected SMGs have this huge apparent CO-luminosity. Our work demonstrates that all-sky surveys are indispensable to find the brightest sources in the universe. In order to do a detailed study of the Cosmic Eyebrow including a lens model for the cold ISM, higher resolution, subarcsecond imaging with ALMA is needed, see e.g., \citet{dye15} and \citet{swi15}. In addition, high-J CO line observations are indispensable to derive molecular gas properties such as gas density and excitation temperature.

%==========
%Table - Property
\begin{table*}
\begin{center}
\caption{Properties of the Cosmic Eyebrow\label{tab:prop}.}
\begin{tabular}{lccc}
\hline\hline
Property&Unit&CO32-A&CO32-B\\
\hline
z$_\mathrm{CO(3-2)}$&&$2.04006\pm0.00003$&$2.0406\pm0.00007$\\
z$_\mathrm{H\alpha}$&&$2.0439\pm0.0006$&...\\
magnification factor $\mu$&&$11\pm$2&$15\pm$3\\
$\mu$L$^{\prime}_\mathrm{CO(1-0)}$&$\times~10^{10}~K~km~s^{-1}~pc^{2}$&$121.6\pm$30.0&...\\
$\mu$L$^{\prime}_\mathrm{CO(3-2)}$&$\times~10^{10}~K~km~s^{-1}~pc^{2}$&$123.8\pm$2.2&$37.3\pm 1.6~$\\
$\mu$L$^{\prime}_\mathrm{[CI(1-0)]}$&$\times~10^{10}~K~km~s^{-1}~pc^{2}$&$27.0\pm$4.0&$...$\\
$\mu$L$_\mathrm{IR}$&$\times~10^{13}~L_{\sun}$&$12.6\pm 1.1~$&$4.5\pm 0.4~$\\
$\mu$SFR$_\mathrm{IR}$$^{a}$&M$_{\sun}$~yr$^{-1}$&$21700$&$8125$\\
$\mu$M$_\mathrm{mol}$$^{b}$, \scriptsize{based on CO(1-0)}&$\times$~10$^{11}$~M$_{\sun}$& 13.2$\pm 3.3$&4.1$^{c}$$\pm 0.1$\\
$\mu$M$_\mathrm{mol}$, \scriptsize{based on dust measurements}&$\times$~10$^{11}$~M$_{\sun}$&$15.1-22.8$&$\approx6.0$\\
M$_\mathrm{mol}$/M$_\mathrm{\ast}$$^{d}$&&0.44&...\\
SFE$_\mathrm{CO(1-0)}$&$L_{\sun}/(K~km~s^{-1}~pc^{2}$)&$103$&...\\
SFE$_\mathrm{CO(3-2)}$&$L_{\sun}/(K~km~s^{-1}~pc^{2}$)&$102$&120\\
$t_{dep}$&Myr&$60$&$50$\\
\hline
\end{tabular}
\label{tab:prop}
\end{center}
\tablecomments{a: The SFR$_{IR}$ is based on \citet{ken98}. b: The molecular gas mass estimate takes into account Helium. c: We used the CO(3-2) observations for component CO32-B and assumed a constant brightness temperature. d: The molecular gas mass estimates is based on the CO(1-0) measurement.}
\end{table*}

\acknowledgments
This work is based on observations carried out under project number D17AA with the IRAM NOEMA interferometer and 170-17 with the IRAM~30m telescope. The Green Bank Observatory is a facility of the National Science Foundation operated under cooperative agreement by Associated Universities, Inc. IRAM is supported by INSU/CNRS (France), MPG (Germany) and IGN (Spain). We thank the anonymous referee for her or his comments that helped us to improve our arguments and presentation in this paper. HD acknowledges financial support from the Spanish Ministry of Economy and Competitiveness (MINECO) under the 2014 Ram\'{o}n y Cajal program MINECO RYC-2014-15686. KCH acknowledges support from the Collaborative Research Center 956, subproject A1, funded by the Deutsche Forschungsgemeinschaft (DFG). KCH would like to thank all of the observers, operators and staff at the GBT and IRAM 30m. We thank Bjorn Emonts for providing the table with literature values for Fig.~\ref{fig:ism}. This research was supported by the Munich Institute for Astro- and Particle Physics (MIAPP) of the DFG cluster of excellence "Origin and Structure of the Universe". This work is carried out within the Collaborative Research Centre 956, sub-project [A1], funded by the Deutsche Forschungsgemeinschaft (DFG). This work has been partially funded by project Spanish Space  Research Programme 'Participation in the NISP instrument and preparation for the science of EUCLID' (ESP2017-84272-C2-1-R and ESP2017-84272-C2-2-R) and AYA2015-69350-C3-3-P (MINECO). This publication makes use of data products from the Wide-field Infrared Survey Explorer, which is a joint project of the University of California, Los Angeles, and the Jet Propulsion Laboratory/California Institute of Technology, funded by the National Aeronautics and Space Administration.


\begin{thebibliography}{}
\bibitem[Alaghband-Zadeh et al.(2012)]{ala12} Alaghband-Zadeh, S., Chapman, S.~C., Swinbank, A.~M., et al.\ 2012, \mnras, 424, 2232

\bibitem[Alaghband-Zadeh et al.(2013)]{ala13} Alaghband-Zadeh, S., Chapman, S.~C., Swinbank, A.~M., et al.\ 2013, \mnras, 435, 1493 

\bibitem[Arabsalmani et al.(2018)]{ara18} Arabsalmani, M., Le Floc'h, E., Dannerbauer, H., et al.\ 2018, \mnras, 476, 2332 

\bibitem[Aravena et al.(2016)]{ara16} Aravena, M., Spilker, J.~S., Bethermin, M., et al.\ 2016, \mnras, 457, 4406 

\bibitem[Becker et al.(1994)]{bec94} Becker, R.~H., White, R.~L., \& Helfand, D.~J.\ 1994, Astronomical Data Analysis Software and Systems III, 61, 165 

\bibitem[B{\'e}thermin et al.(2015)]{bet15} B{\'e}thermin, M., De Breuck, C., Sargent, M., \& Daddi, E.\ 2015, \aap, 576, L9 

\bibitem[Blain(1996)]{bla96} Blain, A.~W.\ 1996, \mnras, 283, 1340 

\bibitem[Bothwell et al.(2013)]{bot13} Bothwell, M.~S., Smail, I., Chapman, S.~C., et al.\ 2013, \mnras, 429, 3047 

\bibitem[Bothwell et al.(2017)]{bot17} Bothwell, M.~S., Aguirre, J.~E., Aravena, M., et al.\ 2017, \mnras, 466, 2825 

\bibitem[Bournaud et al.(2014)]{bou14} Bournaud, F., Perret, V., Renaud, F., et al.\ 2014, \apj, 780, 57 

\bibitem[Calistro-Rivera et al.(2018)]{cal18} Calistro Rivera, G., Hodge, J.~A., Smail, I., et al.\ 2018, \apj, 863, 56 

\bibitem[Ca{\~n}ameras et al.(2015)]{can15} Ca{\~n}ameras, R., Nesvadba, N.~P.~H., Guery, D., et al.\ 2015, \aap, 581, A105 

\bibitem[Carilli \& Walter(2013)]{car13} Carilli, C.~L., \& Walter, F.\ 2013, \araa, 51, 105 

\bibitem[Carlstrom et al.(2011)]{car11} Carlstrom, J.~E., Ade, P.~A.~R., Aird, K.~A., et al.\ 2011, \pasp, 123, 568 

\bibitem[Carter et al.(2012)]{car12} Carter, M., Lazareff, B., Maier, D., et al.\ 2012, \aap, 538, A89 

\bibitem[Casey et al.(2014)]{cas14} Casey, C.~M., Narayanan, D., \& Cooray, A.\ 2014, \physrep, 541, 45 

\bibitem[Chen et al.(2017)]{che17} Chen, C.-C., Hodge, J.~A., Smail, I., et al.\ 2017, \apj, 846, 108 
 
\bibitem[Cox et al.(2011)]{cox11} Cox, P., Krips, M., Neri, R., et al.\ 2011, \apj, 740, 63 

\bibitem[da Cunha et al.(2012)]{dac12} da Cunha, E., Charlot, S., Dunne, L., Smith, D., \& Rowlands, K.\ 2012, The Spectral Energy Distribution of Galaxies - SED 2011, 284, 292 

\bibitem[Daddi et al.(2004)]{dad04} Daddi, E., Cimatti, A., Renzini, A., et al.\ 2004, \apj, 617, 746 

\bibitem[Daddi et al.(2015)]{dad15} Daddi, E., Dannerbauer, H., Liu, D., et al.\ 2015, A\&A, 577, 46

\bibitem[Danielson et al.(2011)]{dan11} Danielson, A.~L.~R., Swinbank, A.~M., Smail, I., et al.\ 2011, \mnras, 410, 1687 

\bibitem[Dannerbauer et al.(2009)]{dan09} Dannerbauer, H., Daddi, E., Riechers, D.~A., et al.\ 2009, \apjl, 698, L178 

\bibitem[Dannerbauer et al.(2017)]{dan17} Dannerbauer, H., Lehnert, M.~D., Emonts, B., et al.\ 2017, \aap, 608, A48 

\bibitem[D{\'{\i}}az-S{\'a}nchez et al.(2017)]{dia17} D{\'{\i}}az-S{\'a}nchez, A., Iglesias-Groth, S., Rebolo, R., \& Dannerbauer, H.\ 2017, \apjl, 843, L22 

\bibitem[Dye et al.(2015)]{dye15} Dye, S., Furlanetto, C., Swinbank, A.~M., et al.\ 2015, \mnras, 452, 2258 

\bibitem[Eales et al.(2010)]{eal10} Eales, S., Dunne, L., Clements, D., et al.\ 2010, \pasp, 122, 499 

\bibitem[Emonts et al.(2016)]{emo16} Emonts, B.~H.~C., Lehnert, M.~D., Villar-Mart{\'{\i}}n, M., et al.\ 2016, Science, 354, 1128 

\bibitem[Emonts et al.(2018)]{emo18} Emonts, B.~H.~C., Lehnert, M.~D., Dannerbauer, H., et al.\ 2018, \mnras, 477, L60 

\bibitem[Engel et al.(2010)]{eng10} Engel, H., Tacconi, L.~J., Davies, R.~I., et al.\ 2010, \apj, 724, 233 

\bibitem[Fixsen et al.(1996)]{fix96} Fixsen, D.~J., Cheng, E.~S., Gales, J.~M., et al.\ 1996, \apj, 473, 576 

\bibitem[Fixsen et al.(1999)]{fix99} Fixsen, D.~J., Bennett, C.~L., \& Mather, J.~C.\ 1999, \apj, 526, 207 

\bibitem[Flower et al.(1994)]{flo94} Flower, D.~R., Le Bourlot, J., Pineau Des Forets, G., \& Roueff, E.\ 1994, \aap, 282, 225 

\bibitem[Gerin \& Phillips(2000)]{ger00} Gerin, M., \& Phillips, T.~G.\ 2000, \apj, 537, 644 

\bibitem[Harrington et al.(2016)]{har16} Harrington, K.~C., Yun, M.~S., Cybulski, R., et al.\ 2016, \mnras, 458, 4383 

\bibitem[Harrington et al.(2018)]{har18} Harrington, K.~C., Yun, M.~S., Magnelli, B., et al.\ 2018, \mnras, 474, 3866 

\bibitem[Harris et al.(2012)]{har12} Harris, A.~I., Baker, A.~J., Frayer, D.~T., et al.\ 2012, \apj, 752, 152 

\bibitem[Hodge et al.(2013)]{hod13} Hodge, J.~A., Karim, A., Smail, I., et al.\ 2013, \apj, 768, 91 

\bibitem[Karim et al.(2013)]{kar13} Karim, A., Swinbank, A.~M., Hodge, J.~A., et al.\ 2013, \mnras, 432, 2 

\bibitem[Iglesias-Groth et al.(2017)]{igl17} Iglesias-Groth, S., D{\'{\i}}az-S{\'a}nchez, A., Rebolo, R., \& Dannerbauer, H.\ 2017, \mnras, 467, 330 

\bibitem[Israel et al.(2015)]{isr15} Israel, F.~P., Rosenberg, M.~J.~F., \& van der Werf, P.\ 2015, \aap, 578, A95 

\bibitem[Ivison et al.(2010)]{ivi10} Ivison, R.~J., Swinbank, A.~M., Swinyard, B., et al.\ 2010, \aap, 518, L35 

\bibitem[Jones (2015)]{jon15} Jones, S. F. 2015, PhD thesis, Univ. Leicester, https://core.ac.uk/download/
pdf/42018134.pdf

\bibitem[Jullo et al.(2007)]{jul07} Jullo, E., Kneib, J.-P., Limousin, M., et al.\ 2007, New Journal of Physics, 9, 447 

\bibitem[Kamenetzky et al.(2016)]{kam16} Kamenetzky, J., Rangwala, N., Glenn, J., Maloney, P.~R., \& Conley, A.\ 2016, \apj, 829, 93 

\bibitem[Kennicutt(1998)]{ken98} Kennicutt, R.~C., Jr.\ 1998, \araa, 36, 189 

\bibitem[Kneib et al.(1993)]{kne93} Kneib, J.~P., Mellier, Y., Fort, B., \& Mathez, G.\ 1993, \aap, 273, 367 

\bibitem[Lestrade et al.(2010)]{les10} Lestrade, J.-F., Combes, F., Salom{\'e}, P., et al.\ 2010, \aap, 522, L4 

\bibitem[Lestrade et al.(2011)]{les11} Lestrade, J.-F., Carilli, C.~L., Thanjavur, K., et al.\ 2011, \apjl, 739, L30 

\bibitem[Liu et al.(2017)]{liu17} Liu, L., Wei{\ss}, A., Perez-Beaupuits, J.~P., et al.\ 2017, \apj, 846, 5 

\bibitem[Madau \& Dickinson(2014)]{mad14} Madau, P., \& Dickinson, M.\ 2014, \araa, 52, 415 

\bibitem[Magdis et al.(2014)]{mag14} Magdis, G.~E., Rigopoulou, D., Hopwood, R., et al.\ 2014, \apj, 796, 63 

\bibitem[Marganian et al.(2013)]{mar13} Marganian, P., Garwood, R.~W., Braatz, J.~A., Radziwill, N.~M., \& Maddalena, R.~J.\ 2013, Astrophysics Source Code Library, ascl:1303.019 

\bibitem[McMahon et al.(2013)]{mcm13} McMahon, R.~G., Banerji, M., Gonzalez, E., et al.\ 2013, The Messenger, 154, 35 

\bibitem[Narayanan \& Krumholz(2014)]{nar14} Narayanan, D., \& Krumholz, M.~R.\ 2014, \mnras, 442, 1411 

\bibitem[Negrello et al.(2007)]{neg07} Negrello, M., Perrotta, F., Gonz{\'a}lez-Nuevo, J., et al.\ 2007, \mnras, 377, 1557 

\bibitem[Negrello et al.(2010)]{neg10} Negrello, M., Hopwood, R., De Zotti, G., et al.\ 2010, Science, 330, 800 

\bibitem[Oguri et al.(2012)]{ogu12} Oguri, M., Bayliss, M.~B., Dahle, H., et al.\ 2012, \mnras, 420, 3213 

\bibitem[Oliver et al.(2012)]{oli12} Oliver, S.~J., Bock, J., Altieri, B., et al.\ 2012, \mnras, 424, 1614 

\bibitem[Papadopoulos \& Greve(2004)]{pap04} Papadopoulos, P.~P., \& Greve, T.~R.\ 2004, \apjl, 615, L29 

\bibitem[Papadopoulos et al.(2012)]{pap12} Papadopoulos, P.~P., van der Werf, P.~P., Xilouris, E.~M., et al.\ 2012, \mnras, 426, 2601 

\bibitem[Planck Collaboration et al.(2011)]{pla11} Planck Collaboration, Ade, P.~A.~R., Aghanim, N., et al.\ 2011, \aap, 536, A1 

\bibitem[Planck Collaboration et al.(2014)]{pla14} Planck Collaboration, Ade, P.~A.~R., Aghanim, N., et al.\ 2014, \aap, 571, A16 

\bibitem[Pilbratt et al.(2010)]{pil10} Pilbratt, G.~L., Riedinger, J.~R., Passvogel, T., et al.\ 2010, \aap, 518, L1 

\bibitem[Riechers(2013)]{rie13} Riechers, D.~A.\ 2013, \nat, 502, 459 

\bibitem[Rosenberg et al.(2015)]{ros15} Rosenberg, M.~J.~F., van der Werf, P.~P., Aalto, S., et al.\ 2015, \apj, 801, 72 

\bibitem[Scoville et al.(2014)]{sco14} Scoville, N., Aussel, H., Sheth, K., et al.\ 2014, \apj, 783, 84 

\bibitem[Scoville et al.(2016)]{sco16} Scoville, N., Sheth, K., Aussel, H., et al.\ 2016, \apj, 820, 83 

\bibitem[Scoville et al.(2017)]{sco17} Scoville, N., Lee, N., Vanden Bout, P., et al.\ 2017, \apj, 837, 150 

\bibitem[Serjeant(2012)]{ser12} Serjeant, S.\ 2012, \mnras, 424, 2429 

\bibitem[Sharon et al.(2016)]{sha16} Sharon, C.~E., Riechers, D.~A., Hodge, J., et al.\ 2016, \apj, 827, 18 

\bibitem[Solomon \& Vanden Bout(2005)]{sol05} Solomon, P.~M., \& Vanden Bout, P.~A.\ 2005, \araa, 43, 677 

\bibitem[Spilker et al.(2014)]{spi14} Spilker, J.~S., Marrone, D.~P., Aguirre, J.~E., et al.\ 2014, \apj, 785, 149 

\bibitem[Strandet et al.(2017)]{str17} Strandet, M.~L., Weiss, A., De Breuck, C., et al.\ 2017, \apjl, 842, L15 

\bibitem[Swinbank et al.(2004)]{swi04} Swinbank, A.~M., Smail, I., Chapman, S.~C., et al.\ 2004, \apj, 617, 64

\bibitem[Swinbank et al.(2010)]{swi10} Swinbank, A.~M., Smail, I., Longmore, S., et al.\ 2010, \nat, 464, 733 

\bibitem[Swinbank et al.(2015)]{swi15} Swinbank, A.~M., Dye, S., Nightingale, J.~W., et al.\ 2015, \apjl, 806, L17 

\bibitem[Tauber et al.(2010)]{tau10} Tauber, J.~A., Mandolesi, N., Puget, J.-L., et al.\ 2010, \aap, 520, A1 

\bibitem[Timmons et al.(2015)]{tim15} Timmons, N., Cooray, A., Nayyeri, H., et al.\ 2015, \apj, 805, 140 

\bibitem[Vieira et al.(2013)]{vie13} Vieira, J.~D., Marrone, D.~P., Chapman, S.~C., et al.\ 2013, \nat, 495, 344 

\bibitem[Yang et al.(2017)]{yan17} Yang, C., Omont, A., Beelen, A., et al.\ 2017, \aap, 608, A144 

\bibitem[Walter et al.(2011)]{wal11} Walter, F., Wei{\ss}, A., Downes, D., Decarli, R., \& Henkel, C.\ 2011, \apj, 730, 18 

\bibitem[Wei{\ss} et al.(2003)]{wei03} Wei{\ss}, A., Henkel, C., Downes, D., \& Walter, F.\ 2003, \aap, 409, L41 

\bibitem[Wei{\ss} et al.(2005)]{wei05} Wei{\ss}, A., Downes, D., Henkel, C., \& Walter, F.\ 2005, \aap, 429, L25 


\bibitem[Whitaker et al.(2012)]{whi12} Whitaker, K.~E., van Dokkum, P.~G., Brammer, G., \& Franx, M.\ 2012, \apjl, 754, L29 

\bibitem[Wright et al.(2010)]{wri10} Wright, E.~L., Eisenhardt, P.~R.~M., Mainzer, A.~K., et al.\ 2010, \aj, 140, 1868 

\bibitem[Xie et al.(1995)]{xie95} Xie, T., Allen, M., \& Langer, W.~D.\ 1995, \apj, 440, 674 

\end{thebibliography}
\end{document}